\documentclass[twoside,twocolumn]{article}

\usepackage{blindtext} 

\usepackage[sc]{mathpazo} 
\usepackage[T1]{fontenc} 
\usepackage{microtype} 

\usepackage[english]{babel} 

\usepackage[hmarginratio=1:1,top=32mm,columnsep=20pt]{geometry} 
\usepackage[hang, small,labelfont=bf,up,textfont=it,up]{caption} 
\usepackage{booktabs} 

\usepackage{lettrine} 

\usepackage{enumitem} 
\setlist[itemize]{noitemsep} 

\usepackage{abstract} 

\usepackage{titlesec} 
\renewcommand\thesection{\Roman{section}} 
\renewcommand\thesubsection{\roman{subsection}} 
\titleformat{\section}[block]{\large\scshape\centering}{\thesection.}{1em}{} 
\titleformat{\subsection}[block]{\large}{\thesubsection.}{1em}{} 

\usepackage{fancyhdr} 
\usepackage{titling} 

\usepackage{hyperref} 

\usepackage{scalerel, amssymb}

\def\msquare{\mathord{\scalerel*{\Box}{gX}}}
\usepackage{amsmath}
\usepackage{framed}
\usepackage{graphicx} 
\usepackage{hyperref}   
\usepackage{multicol}
\usepackage{float}
\usepackage{nomencl}
\usepackage{wasysym}
\hypersetup{
	colorlinks=true,
	linkcolor=blue,
	citecolor=blue,
	urlcolor=blue,
}
\usepackage[square,numbers]{natbib}

\newcommand{\KXX}{K_{XX}}
\newcommand{\KXY}{K_{XY}}
\newcommand{\KXA}{K_{X\alpha}}
\newcommand{\KXB}{K_{X\beta}}
\newcommand{\KYX}{K_{YX}}
\newcommand{\KYY}{K_{YY}}
\newcommand{\KYA}{K_{Y\alpha}}
\newcommand{\KYB}{K_{Y\beta}}
\newcommand{\KAX}{K_{\alpha X}}
\newcommand{\KAY}{K_{\alpha Y}}
\newcommand{\KAA}{K_{\alpha \alpha}}
\newcommand{\KAB}{K_{\alpha \beta}}
\newcommand{\KBX}{K_{\beta X}}
\newcommand{\KBY}{K_{\beta Y}}
\newcommand{\KBA}{K_{\beta \alpha}}
\newcommand{\KBB}{K_{\beta \beta}}
\newcommand{\CXX}{C_{XX}}
\newcommand{\CXY}{C_{XY}}
\newcommand{\CXA}{C_{X\alpha}}
\newcommand{\CXB}{C_{X\beta}}
\newcommand{\CYX}{C_{YX}}
\newcommand{\CYY}{C_{YY}}
\newcommand{\CYA}{C_{Y\alpha}}
\newcommand{\CYB}{C_{Y\beta}}
\newcommand{\CAX}{C_{\alpha X}}
\newcommand{\CAY}{C_{\alpha Y}}
\newcommand{\CAA}{C_{\alpha \alpha}}
\newcommand{\CAB}{C_{\alpha \beta}}
\newcommand{\CBX}{C_{\beta X}}
\newcommand{\CBY}{C_{\beta Y}}
\newcommand{\CBA}{C_{\beta \alpha}}
\newcommand{\CBB}{C_{\beta \beta}}
\newcommand{\MXX}{M_{XX}}
\newcommand{\MXY}{M_{XY}}
\newcommand{\MXA}{M_{X\alpha}}
\newcommand{\MXB}{M_{X\beta}}
\newcommand{\MYX}{M_{YX}}
\newcommand{\MYY}{M_{YY}}
\newcommand{\MYA}{M_{Y\alpha}}
\newcommand{\MYB}{M_{Y\beta}}
\newcommand{\MAX}{M_{\alpha X}}
\newcommand{\MAY}{M_{\alpha Y}}
\newcommand{\MAA}{M_{\alpha \alpha}}
\newcommand{\MAB}{M_{\alpha \beta}}
\newcommand{\MBX}{M_{\beta X}}
\newcommand{\MBY}{M_{\beta Y}}
\newcommand{\MBA}{M_{\beta \alpha}}
\newcommand{\MBB}{M_{\beta \beta}}

\pretitle{\begin{center}\huge\bfseries} 
\posttitle{\end{center}} 
\title{Dynamic force and torque characteristic \\ of annular gaps -
Simulation results and evaluation of the relevance of the tilt and torque coefficients} 
\author{
\textsc{Maximilian M. G. Kuhr}\thanks{Corresponding author} \\[1ex]
\normalsize Chair of Fluid Systems \\
\normalsize Technische Universität Darmstadt \\
\normalsize \href{mailto:maximilian.kuhr@fst.tu-darmstadt.de}{maximilian.kuhr@fst.tu-darmstadt.de}
\and 
\textsc{Rainer Nordmann} \\[1ex]
\normalsize Fraunhofer Institute for \\
\normalsize Structural Durability and System Reliability \\
\normalsize \href{mailto:rainer.nordmann@lbf-extern.fraunhofer.de}{rainer.nordmann@lbf-extern.fraunhofer.de}
\and 
\textsc{Peter F. Pelz} \\[1ex]
\normalsize Chair of Fluid Systems \\
\normalsize Technische Universität Darmstadt \\
\normalsize \href{mailto:peter.pelz@fst.tu-darmstadt.de}{peter.pelz@fst.tu-darmstadt.de}
}
\date{} 
\renewcommand{\maketitlehookd}{%
\begin{abstract}
We examine the dynamic force and torque characteristic of annular gaps resulting from an axial flow component. First, the rotordynamic influence of annular gaps with an axial flow component is recapitulated. Until now, there is a major deficit of understanding the dynamic force and torque characteristic of annular gaps. Second, a new simulation method is presented, using a perturbed integro-differential approach in combination with a Hirs' model and power-law ansatz functions for the velocity profiles to calculate the dynamic force and torque characteristics. Third, an extensive parameter study is carried out. To evaluate whether or not the hydraulic tilt and torque coefficients need to be taken into account, an effective stiffness is defined and the influence of the annulus length, an eccentrically operated shaft, the centre of rotation, a modified Reynolds number, the flow number and the pre-swirl is investigated.  
\end{abstract}
}

\begin{document}

\maketitle


\section{Introduction}
The reliability and performance of turbomachinery is often limited by harmful shaft vibrations due to resonance effects or the response of the system to disturbances during stationary operation. The dynamic behaviour of the system and therefore the mechanical vibrations are highly influenced by the induced hydrodynamic forces and torques of the flow within narrow annular gaps \citep{Childs.1983, Childs.1983b,Childs.1993, SanAndres.1991b,SanAndres.1993, SanAndres.1993b, SanAndres.1993c,Tiwari.2018}. In general, the flow in an annulus is three-dimensional. The presence of an axial pressure difference results in an axial flow component that is superimposed by the circumferential flow component driven by viscous forces. In addition, the flow at the annulus inlet is superimposed by a pre-swirl due to the design parameters of the turbomachinery. The pre-swirl is then convected into the annuls by the axial flow component. Due to the increasing demands on flexibility, today's turbomachinery are often operated at partial load under highly dynamic operating conditions. Therefore, increased vibrations occur due to flow separation and recirculation areas, which pose a challenge to service life and safe operation.
By now there is a major deficit in understanding the dynamic characteristic of the induced hydraulic forces and torques in annular gaps. Furthermore, the existing literature mainly focuses on the influence of hydraulic forces due to translational motion whereas the influence of hydraulic tilt and torque coefficients and their influence on the stability of the rotor-annulus systems are invariably neglected. In modern turbomachinery, two essential narrow annular gaps exist, applying hydrodynamic forces and torques on the rotating shaft \citep{Childs.1993, Gasch.2002, Gulich.2010}. First, annular seals or damper seals and second, journal bearings which are either oil or media lubricated. However, the usage of low viscous fluids like water or cryogenic liquids for lubrication purpose, as is usual for annular seals and media lubricated bearings, leads to an operation at high Reynolds numbers, resulting in turbulent flow conditions and significant inertia effects \citep{Childs.1993, SanAndres.1991b,SanAndres.1993, SanAndres.1993b, SanAndres.1993c}. The rotordynamic influence of those annuli is in general described by the use of rotordynamic coefficients, namely stiffness $\tilde{K}$, damping $\tilde{C}$ and inertia $\tilde{M}$. Here, the tilde $\tilde{\msquare}$ characterises dimensional variables. The generalised equation of motion including forces and torques of the annular gap flow yields
\begin{equation}\label{eqn:introduction_full_matrix}
	\begin{split}
		-\begin{bmatrix}
			\tilde{F}_{X}\\
			\tilde{F}_{Y}\\
			\tilde{M}_{X}\\
			\tilde{M}_{Y}
		\end{bmatrix} = & \begin{bmatrix}
			\tilde{K}_{XX} &  \tilde{K}_{XY} & \tilde{K}_{X\alpha} & \tilde{K}_{X \beta}\\
			\tilde{K}_{YX} &  \tilde{K}_{YY} & \tilde{K}_{Y\alpha} & \tilde{K}_{Y \beta}\\
			\tilde{K}_{\alpha X} &  \tilde{K}_{\alpha Y} & \tilde{K}_{\alpha\alpha} & \tilde{K}_{\alpha\beta}\\
			\tilde{K}_{\beta X} &  \tilde{K}_{\beta Y} & \tilde{K}_{\beta\alpha} & \tilde{K}_{\beta\beta}
		\end{bmatrix} \begin{bmatrix} 
			\tilde{X}\\
			\tilde{Y}\\
			\alpha_X\\
			\beta_Y
		\end{bmatrix} \, +\\  &\, + \begin{bmatrix}
			\tilde{C}_{XX} &  \tilde{C}_{XY} & \tilde{C}_{X\alpha} & \tilde{C}_{X \beta}\\
			\tilde{C}_{YX} &  \tilde{C}_{YY} & \tilde{C}_{Y\alpha} & \tilde{C}_{Y \beta}\\
			\tilde{C}_{\alpha X} &  \tilde{C}_{\alpha Y} & \tilde{C}_{\alpha\alpha} & \tilde{C}_{\alpha\beta}\\
			\tilde{C}_{\beta X} &  \tilde{C}_{\beta Y} & \tilde{C}_{\beta\alpha} & \tilde{C}_{\beta\beta}
		\end{bmatrix} \begin{bmatrix} 
			\tilde{\dot{X}}\\
			\tilde{\dot{Y}}\\
			\dot{\alpha}_X\\
			\dot{\beta}_Y
		\end{bmatrix} \, +\\  &\quad + \begin{bmatrix}
			\tilde{M}_{XX} &  \tilde{M}_{XY} & \tilde{M}_{X\alpha} & \tilde{M}_{X \beta}\\
			\tilde{M}_{YX} &  \tilde{M}_{YY} & \tilde{M}_{Y\alpha} & \tilde{M}_{Y \beta}\\
			\tilde{M}_{\alpha X} &  \tilde{M}_{\alpha Y} & \tilde{M}_{\alpha\alpha} & \tilde{M}_{\alpha\beta}\\
			\tilde{M}_{\beta X} &  \tilde{M}_{\beta Y} & \tilde{M}_{\beta\alpha} & \tilde{M}_{\beta\beta}
		\end{bmatrix} \begin{bmatrix} 
			\tilde{\ddot{X}}\\
			\tilde{\ddot{Y}}\\
			\ddot{\alpha}_X\\
			\ddot{\beta}_Y
		\end{bmatrix}.
	\end{split}
\end{equation} 

\begin{figure*}
	\centering
	\includegraphics[scale=1.0]{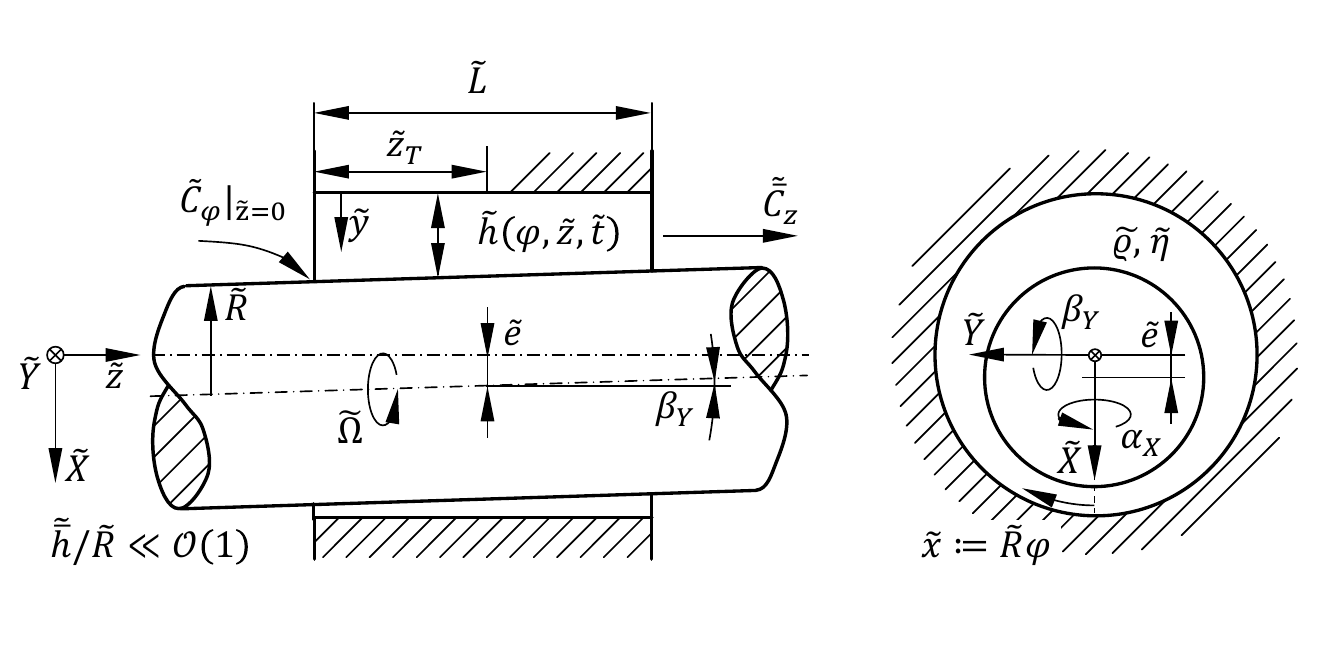}
	\caption{Schematic drawing of an eccentrically operated generic annular gap with axial flow and pre-swirl at the annulus inlet.}
	\label{fig:introduction_generic_annulus}
\end{figure*}

Here, $\tilde{F}_X, \tilde{F}_Y$ and $\tilde{M}_X, \tilde{M}_Y$ are the induced hydrodynamic forces and torques of the annulus acting on the rotor. $\tilde{X}, \tilde{\dot{X}}, \tilde{\ddot{X}}$ and $\tilde{Y},\tilde{\dot{Y}},\tilde{\ddot{Y}}$ denotes the translational motion of the rotor and its time derivatives. $\alpha_X, \dot{\alpha}_X, \ddot{\alpha}_X$ and $\beta_Y, \dot{\beta}_Y, \ddot{\beta}_Y$ is the angular motion of the rotor around the $\tilde{X}, \tilde{Y}$ axis and its time derivatives. The $48$ rotordynamic coefficients are in general dependent on (i) the geometry of the annulus, i.e. the mean gap height $\tilde{\bar{h}}$, the shaft radius $\tilde{R}$, the gap length $\tilde{L}$ and the gap function  $\tilde{h} = \tilde{h}\,(\varphi, \tilde{z}, \tilde{t})$ with the circumferential and axial coordinates $\varphi, \tilde{z}$; (ii) the operating parameters of the system, i.e. the static eccentric and angular position of the shaft inside the annulus $\tilde{e}, \alpha, \beta$, the distance of the fulcrum $\tilde{z}_T$ from the gap entrance, the angular velocity of the shaft $\tilde{\Omega}$, the mean axial velocity through the annulus $\tilde{\bar{C}}_z$ and the pre-swirl velocity at the gap entrance $\tilde{C}_\varphi|_{z=0}$; (iii) the characteristics of the fluid used for lubrication, i.e. fluid density $\tilde{\varrho}$ and the dynamic viscosity $\tilde{\eta}$, cf. figure, \ref{fig:introduction_generic_annulus}
\begin{equation}\label{eqn:dependance_coefficients}
	\begin{split}
		&\tilde{K}_{ij}, \tilde{C}_{ij}, \tilde{M}_{ij} = \\
		&=f\left(\tilde{R}, \tilde{\bar{h}}, \tilde{L}, \tilde{e}, \alpha_X, \beta_Y, \tilde{z}_T, \tilde{\Omega}, \tilde{\bar{C}}_z, \tilde{C}_\varphi|_{z=0}, \tilde{\varrho}, \tilde{\eta}\right).
	\end{split}
\end{equation}

On dimensional ground, the dimensionless rotordynamic coefficients are only a function of 9 dimensionless measures: (i) the relative gap clearance $\psi:=\tilde{\bar{h}}/\tilde{R}$, (ii) the dimensionless annulus length $L:=\tilde{L}/\tilde{R}$, (iii) the relative eccentricity $\varepsilon:=\tilde{e}/\tilde{\bar{h}}$, (iv) the dimensionless fulcrum $z_T:=\tilde{z}_T/\tilde{L}$, (v) the Reynolds number in circumferential direction $Re_\varphi:= (\tilde{\Omega}\tilde{R}\tilde{\bar{h}}/\tilde{\nu})$, (vi) the flow number $\phi:=\tilde{\bar{C}}_z/(\tilde{\Omega}\tilde{R})$, (vii) the dimensionless pre-swirl $C_\varphi\vert_{z=0}:=\tilde{C}_\varphi/(\tilde{\Omega}\tilde{R})$ and (viii + ix) the angular displacements $\alpha:=\tilde{L}\alpha_X/\tilde{\bar{h}}$, $\beta:=\tilde{L}\beta_Y/\tilde{\bar{h}}$ around the fulcrum. Kuhr et al. \citep{Kuhr.Tobepublishedb} showed that the list of dimensionless measures can be further reduced because the relative gap clearance and Reynolds number only appear as a product in the governing equations, resulting in a new dimensionless measure, the modified Reynolds number $Re_\varphi^*:=\psi Re^{n_f}_\varphi$. Here $n_f$ is an empirical constant describing an arbitrary line within the double logarithmic Moody diagram. Hence, equation~\ref{eqn:dependance_coefficients} is reduced to
\begin{equation}
	\begin{split}
		K_{ij}, C_{ij}&, M_{ij} = \\
		&=f\left(L, \varepsilon, z_T, Re^*_\varphi, \phi, C_\varphi\vert_{z=0}, \alpha, \beta\right).
	\end{split}
\end{equation}

The dimensionless rotordynamic coefficients are defined for the force and torque due to translational and angular motion separately
\begin{equation}
	\begin{split}
		K_{I} := \frac{2\tilde{\bar{h}}\tilde{K}_{I}}{\tilde{\varrho}\tilde{\Omega}^2\tilde{R}^3\tilde{L}}, \quad K_{II} := \frac{2\tilde{\bar{h}}\tilde{K}_{II}}{\tilde{\varrho}\tilde{\Omega}^2\tilde{R}^3\tilde{L}^2},\\
		K_{III} := \frac{2\tilde{\bar{h}}\tilde{K}_{III}}{\tilde{\varrho}\tilde{\Omega}^2\tilde{R}^3\tilde{L}^2}, \quad K_{IV} := \frac{2\tilde{\bar{h}}\tilde{K}_{IV}}{\tilde{\varrho}\tilde{\Omega}^2\tilde{R}^3\tilde{L}^3}.
	\end{split}
\end{equation}

Here, the first two stiffness coefficients represent dynamic force coefficients due to translational and angular motion whereas the two later ones represent the torque coefficients due to translational and angular motion. The indices represent the corresponding sub-matrices of equation \ref{eqn:introduction_full_matrix}. The damping and inertia terms are defined accordingly
\begin{equation}
	\begin{split}
		C_{I} := \frac{2\tilde{\bar{h}}\tilde{C}_{I}}{\tilde{\varrho}\tilde{\Omega}\tilde{R}^3\tilde{L}}, \quad C_{II} := \frac{2\tilde{\bar{h}}\tilde{C}_{II}}{\tilde{\varrho}\tilde{\Omega}\tilde{R}^3\tilde{L}^2},\\
		C_{III} := \frac{2\tilde{\bar{h}}\tilde{C}_{III}}{\tilde{\varrho}\tilde{\Omega}\tilde{R}^3\tilde{L}^2}, \quad C_{IV} := \frac{2\tilde{\bar{h}}\tilde{C}_{IV}}{\tilde{\varrho}\tilde{\Omega}\tilde{R}^3\tilde{L}^3},\\
		M_{I} := \frac{2\tilde{\bar{h}}\tilde{M}_{I}}{\tilde{\varrho}\tilde{R}^3\tilde{L}}, \quad M_{II} := \frac{2\tilde{\bar{h}}\tilde{M}_{II}}{\tilde{\varrho}\tilde{R}^3\tilde{L}^2},\\
		M_{III} := \frac{2\tilde{\bar{h}}\tilde{M}_{III}}{\tilde{\varrho}\tilde{R}^3\tilde{L}^2}, \quad M_{IV} := \frac{2\tilde{\bar{h}}\tilde{M}_{IV}}{\tilde{\varrho}\tilde{R}^3\tilde{L}^3}.
	\end{split}
\end{equation}

As mentioned earlier, the vast majority of the existing literature, whether the studies are of analytical, experimental or numerical nature, focuses only on the rotordynamic influence of induced forces due to translational but not angular motion. Furthermore, either annular seals operating at zero eccentricity with high axial pressure differences, cf. \citep{AlQutub.2000, Amoser.1995, Arghir.2001, Arghir.2004, Childs.1983,Childs.1983b, Childs.1990b, Childs.2008, Moreland.2018, Dietzen.1988b, Nordmann.1987, Nordmann.1986, Nordmann.1984, SanAndres.2006, SanAndres.2018} or journal bearings operating at high eccentricities without an axial pressure difference, cf. \citep{Constantinescu.1982, Glienicke.1967, Frene.2006, Feng.2017, Yuan.2006, Feng.2019, Wang.2013, Dousti.2016} are discussed. \\
A more detailed list of the research of annular seals and journal bearings can be found in the work of Tiwari et al. \citep{Tiwari.2005, Tiwari.2004}. \\
With regard to the hydraulic torques, Childs \citep{Childs.1982} presented a bulk-flow based calculation method for determining skew-symmetric dynamic force and torque coefficients for finite length annular pressure seals operating at zero eccentricity under fully turbulent flow conditions. The static and dynamic properties are calculated by linearising the non-linear system of partial differential equations by means of a perturbation analysis, leading to a set of linear ordinary differential equations. Focusing mainly on the effect of the seal length on the coefficients at a constant pressure difference without pre-swirl. Childs examined the coefficients for three different seal lengths $\tilde{L}/\tilde{R}=0.3, 1.0, 2.0$. In addition, one calculation with a negative pre-swirl ratio is carried out at a seal length of $\tilde{L}/\tilde{R}=1.0$ to show the effect of pre-swirl on the coefficients.\\ 
Simon \& Fr\^{e}ne \citep{Simon.1992} developed an analysis to calculate the static and dynamic characteristics of annular seals at eccentric shaft operation. Similar to the work of Childs \citep{Childs.1982} the method is based on an integro-differential approach, solving the continuity and momentum equation in axial and circumferential direction. In contrast to Childs, the integrals are not solved with the bulk-flow approach but are modelled using parabolic ansatz functions. The governing equations are expanded using a perturbation method, leading to a system of non-linear partial differential equations for the zeroth-order and a linear system of partial differential equations for the first-order solution. Here, the zeroth-order solution gives the static characteristics of the flow whereas the first-order solution gives the dynamic properties. Both equation systems are solved numerically by a shooting method. For the purpose of validation the numeric results are first compared to experimental and numerical data for rotordynamic force coefficients due to translational displacement provided by Nordmann \& Dietzen \citep{Nordmann.1988} and Nelson \& Nguyen \citep{Nelson.1988b}. The results are in good agreement with the chosen validation data. The paper then gives the stiffness and damping coefficients for angular excitation for three relative eccentricities $\varepsilon := \tilde{e} / \tilde{\bar{h}} = 0.0, 0.5, 0.8$. Simon \& Fr\^{e}ne \citep{Simon.1992} extend their research on the influence of the annulus length and the pre-swirl on the rotordynamic coefficients. In addition to positive values of the pre-swirl negative values are also investigated and compared to the limited data published by Childs \citep{Childs.1982}.\\
San Andr\'{e}s \citep{SanAndres.1993} presents a method based on the bulk-flow approach, examining dynamic force and torque coefficients for short annular seals of two different lengths $L:=\tilde{L}/\tilde{R}=0.4, 1.0$ operated at zero eccentricity. San Andres compares his approach to the results of Childs \citep{Childs.1982} and Simon \& Fr\^{e}ne \citep{Simon.1992b}. The method showns a good agreement with the rotordynamic coefficients. The paper then mainly focuses on the influence of the fulcrum $\tilde{z}_T$ on the dynamic properties. It is shown that the coefficients are highly sensitive as to whether the centre of rotation is at the entrance, centre or exit of the gap. The method is later used to further investigate the influence of shaft misalignment on the dynamic properties, cf. \citep{SanAndres.1993b}. \\
Here, the dynamic torque coefficients are determined for a centred annular pressure seal at high degrees of static misalignment. The governing two dimensional equations are expanded using a perturbation method leading to a non-linear partial differential equation system for the zeroth-order and a linear partial differential equation system for the first-order solution. Both equation systems are solved numerically by using a SIMPLEC (Semi-Implicit Method for Pressure Linked Equations-Consistent \citep{Schafer.2006}) algorithm coupled to a Newton-Raphson iterative procedure. In addition, the analysis is enhanced by using accurate analytical expressions for the centred operation. It is shown that the skew-symmetry is not longer valid for large static angles of misalignment which intuitively seems correct, since large angles of misalignment in parts of the annulus produce large static eccentricities. \\
San Andr\'{e}s \citep{SanAndres.1993c} extends his theoretical approach to the effects of journal misalignment on the operation of turbulent flow hydrostatic bearings. The work focuses on the effect of eccentricity and misalignment on the stiffness and damping coefficients due to translational and angular motions as well as the inertia terms for translational displacement. Similar to the work of Simon \& Fr\^{e}ne \citep{Simon.1992} no inertia terms for angular displacements are given. It is concluded that the presence of journal misalignment affects the bearing performance and needs to be taken into account when properly designing turbulent flow hydrostatic bearings. \\
Kanemori \& Iwatsubo \citep{Kanemori.1992} present an experimental study of the dynamic force and torque coefficients for long annular seals $L:=\tilde{L}/\tilde{R}=6.0$. The test rig presented consists of a rotor driven by 2 motors to realise the spinning and whirling motion. The induced forces were measured with attached pietzo-electric load cells. The corresponding torques were then calculated around the fulcrum by using the geometric distance and the measured loads. For the purpose of validation the measured dynamic coefficients were compared to the calculation method developed by Childs \citep{Childs.1982}. It is shown that the dynamic coefficients for stiffness and damping coincide well with the theory. In addition, Kanemori \& Iwatsubo \citep{Kanemori.1992} state that the rotor forward whirl acts as a stabilising force on the rotor.\\
Kanemori \& Iwatsubo \citep{Kanemori.1994} studied the mutual effects of cylindrical and conical whirl on the dynamic fluid forces and torques on a long annular seal experimentally. In addition to the rotordynamic coefficients the paper presents pressure measurements inside the annulus. The paper focuses mainly on the impact of phase difference at whirling motion. It is shown that the phase of the whirl acts either as a stabilising or destabilising force. 
Kanemori \& Iwatsubo \citep{Kanemori.1994b} extend their work to a linear stability analysis using the logarithmic decrement on a real submerged motor pump including hydrodynamic forces and torques when carrying out the stability analysis. The individual contributions of the whirl and concentric motion to the stability of the system is evaluated. \\
Feng \& Jiang \citep{Feng.2017} and Feng et al. \citep{Feng.2019} use the Reynolds' equation of lubrication theory to calculate the stiffness and damping coefficients due to translational and angular motions of a water lubricated hydrodynamic journal bearing. The influence of laminar and turbulent flow conditions is investigated as well as the influence of eccentricity, rotating speed and tilting angle. The  Reynolds' equation is therefore modified by using turbulent correction coefficients based on the work of Fr\^{e}ne \& Arghir \citep{Frene.2006}. Unfortunately the effects of fluid inertia, i.e. the inertia terms $M_{ij}$, are neglected.
\section{Governing equations}
In contrast to the majority of the literature we specialise neither on annular seals nor journal bearings, We rather  develope a method based on the generic annulus geometry under turbulent flow conditions, cf. figure \ref{fig:introduction_generic_annulus}. Hence, our method is applicable to annular seals and journal bearings Kuhr et al. \citep{Kuhr.Tobepublishedb} present the Clearance Averaged Pressure Model (CAPM), a method determining the static characteristics of generic annuli. Similar to the bulk-flow approach, the model uses an integro-differential approach but the velocity integrals are treated by using power law ansatz functions. The model is experimentally validated by a specifically designed test rig using active magnetic bearings. In the following, the model is expanded, using a perturbation expansion to determine the dynamic force and torque characteristics. \\

The dimensionless gap function for an eccentric and misaligned shaft reads
\begin{equation}
	h = 1 - \varepsilon \cos\varphi - \left(z-z_T\right)\frac{L}{\psi}\tan\beta.
\end{equation}

Here, $\varepsilon$ and $\beta$ are time dependant. The time dependant dimensionless continuity and momentum equation in circumferential and axial direction yields
\begin{equation}
	\begin{split}
		&\frac{\partial}{\partial t} h + \frac{\partial}{\partial \varphi} h \int_{0}^{1} c_\varphi \, \mathrm{d}y + \frac{\phi}{L}  \frac{\partial}{\partial z} h \int_{0}^{1} c_z \, \mathrm{d}y = 0, \\
		&\frac{\partial}{\partial t} h \int_{0}^{1} c_\varphi \, \mathrm{d}y + \frac{\partial}{\partial \varphi} h \int_{0}^{1} c^2_\varphi \, \mathrm{d}y \;+ \\ 
		&\quad\quad +\,\frac{\phi}{L} \frac{\partial}{\partial z} h \int_{0}^{1} c_\varphi c_z \, \mathrm{d}y =  -\frac{h}{2} \frac{\partial p}{\partial \varphi} + \frac{1}{2\psi} \tau_{yx}|^1_0,\\
		&\phi \frac{\partial}{\partial t} h \int_{0}^{1} c_z \, \mathrm{d}y + \phi \frac{\partial}{\partial \varphi} h \int_{0}^{1} c_\varphi c_z \, \mathrm{d}y\; + \\
		&\quad\quad  + \, \frac{\phi^2}{L} \frac{\partial}{\partial z} h \int_{0}^{1} c^2_z \, \mathrm{d}y = -\frac{h}{2L} \frac{\partial p}{\partial z} + \frac{1}{2\psi} \tau_{yz}|^1_0.
	\end{split}
\end{equation}

By using power law ansatz functions of the form $c=C(2y)^{1/n}$ with the centreline velocities $C_\varphi, C_z$ at half gap height and the exponents $n_\varphi, n_z$, the integrals for the continuity and momentum equation yields
\begin{equation}
	\begin{split}
		& \int_{0}^{1}c_\varphi\,\mathrm{d}y = \frac{n_\varphi}{n_\varphi + 1} C_\varphi + \frac{1}{2\left(n_\varphi + 1 \right)},\\
		& \int_{0}^{1}c_z\,\mathrm{d}y = \frac{n_z}{n_z + 1} C_z,\\
		& \int_{0}^{1}c^2_\varphi\,\mathrm{d}y = \frac{n_\varphi}{\left(n_\varphi+2\right)\left(n_\varphi+1\right)}C_\varphi + \\ 
		& \quad\quad\quad\quad\quad\quad + \frac{n_\varphi}{n_\varphi +2}C^2_\varphi + \frac{1}{\left(n_\varphi+2\right)\left(n_\varphi+1\right)},\\
		& \int_{0}^{1}c_\varphi c_z\,\mathrm{d}y = \frac{n_\varphi n_z}{n_\varphi n_z + n_\varphi + n_z}C_z C_\varphi + \\
		& \quad\quad\quad\quad\quad\quad + \frac{n_z^2}{2\left(n_\varphi n_z + n_\varphi + n_z\right)\left(n_z + 1\right)}C_z,\\
		& \int_{0}^{1}c^2_z\,\mathrm{d}y = \frac{n_z}{n_z + 2}C^2_z.
	\end{split}
\end{equation}

The wall shear stresses $\tau_{yi}|^1_0 = $ are modelled according to the bulk-flow theory for turbulent film flows using Hirs' \citep{Hirs.1973} approach by means of the Fanning friction factor 
\begin{equation}
	\tau_{yi}|^1_0 = \tau_{yi,S} - \tau_{yi,R}.
\end{equation}

The directional wall shear stresses $\tau_{yi,S}$ and $\tau_{yi,R}$ reads,
\begin{equation}
	\begin{split}
		& \tau_{yi,R} = f_R C_R C_{i,R},\\
		& \tau_{yi,S} = f_S C_S C_{i,S}.
	\end{split}
\end{equation}

with the Fanning friction factor $f_i$  ($i=R,S$), the dimensionless effective relative velocity between the wall (rotor $R$, stator $S$) and the fluid $C_i:=\sqrt{C^2_{\varphi,i} + \phi^2 C^2_{z,i}}, i=R,S$. The components $C_{\varphi,i}$ and $C_{z,i}$ are boundary layer averaged velocities between wall and the corresponding boundary layer thickness $\delta$ assuming fully developed boundary layers throughout the annulus. The boundary layer averaged velocities yields
\begin{equation}
	\begin{split}
		& C_{\varphi, S} := \frac{1}{\delta}\int_{0}^{\delta}c_\varphi\,\mathrm{d}y = \frac{n_\varphi}{n_\varphi + 1} C_\varphi,\\
		& C_{\varphi, R} := \frac{1}{\delta}\int_{0}^{\delta}\left(c_\varphi-1\right)\,\mathrm{d}y = \frac{n_\varphi}{n_\varphi + 1} \left(C_\varphi-1\right),\\
		& C_{z, S} := \frac{1}{\delta}\int_{0}^{\delta}c_z\,\mathrm{d}y = \frac{n_z}{n_z + 1} C_z,\\
		& C_{z, R} := \frac{1}{\delta}\int_{0}^{\delta}c_z\,\mathrm{d}y = \frac{n_z}{n_z + 1} C_z.
	\end{split}
\end{equation}

The Fanning friction factor is given by
\begin{equation}
	f_i= m_f \left(\frac{h}{2}C_i Re_\varphi\right)^{-n_f}.
\end{equation}

Here, the friction factor is modelled using the empirical constants $n_f$ and $m_f$ as well as the corresponding velocity at the rotor and stator $C_i$, ($i = R, S$) and the Reynolds number $Re_\varphi$. Arbitrary lines within the double logarithmic Moody diagram give the empirical constants $n_f$ and $m_f$. 
In order to solve the system of equations, boundary conditions have to be specified. The pressure loss at the gap entrance is modelled by applying Bernoulli's equation. The pressure boundary conditions at the gap inlet reads,
\begin{equation}
	 p|_{z=0} = \Delta p - \left(1+\zeta\right)\left(\phi^2 C^2_z + C^2_\varphi\right),\\
\end{equation}

Here, $\Delta p$ is the overall axial pressure difference, whereas $\zeta$ is the entrance pressure loss coefficient. A Dirichlet boundary condition for the pressure is applied at the annulus exit:
\begin{equation}
	p|_{z=1} = 0.
\end{equation}

In addition, the pre-swirl, i.e. the circumferential velocity at the annulus inlet, is applied. This yields,
\begin{equation}
	C_\varphi|_{z=0} \in \mathbb{R}.
\end{equation}  

\subsection{Perturbation analysis}
For the calculation of the dynamic forces and torque characteristics a perturbation analysis is used. Assuming small harmonic disturbances $\Delta$ around the static equilibrium position, the corresponding variables $C_\varphi, C_z, p$ and $h$ are developed using a first-order perturbation expansion. In the perturbation ansatz, the zeroth-order variables are independent of time while the first-order variables are time dependant 
\begin{equation}
	\begin{split}
		C_\varphi\left(t, \varphi, z\right) &= C_{\varphi,0}\left(\varphi, z\right) + \Delta C_{\varphi,1}\left(t, \varphi, z\right),\\
		C_z\left(t, \varphi, z\right) &= C_{z,0}\left(\varphi, z\right) + \Delta C_{z,1}\left(t, \varphi, z\right),\\
		p\left(t, \varphi, z\right) &= p_0\left(\varphi, z\right) + \Delta p_1\left(t, \varphi, z\right),\\
		h\left(t, \varphi, z\right) &= h_0\left(\varphi, z\right) + \Delta h_1\left(t, \varphi, z\right).
	\end{split}
\end{equation}

By inserting the perturbation expansion into the equations as well as into the boundary conditions, a set of partial differential equations for the zeroth- and first-order is derived. The solution of the zeroth-order equation system yields the static characteristics of the annulus, i.e. the static induced forces and torques and the corresponding attitude angle, whereas the solution of the first-order equation system gives the dynamic properties of the system. The zeroth-order equations read
\begin{equation}
	\begin{split}
		&\frac{\partial}{\partial \varphi}h_0\int_{0}^{1}c_{\varphi,0}\,\mathrm{d}y + \frac{\phi}{L} \frac{\partial}{\partial z}h_0\int_{0}^{1}c_{z,0}\,\mathrm{d}y = 0, \\
		&\frac{\partial}{\partial \varphi}h_0\int_{0}^{1}c^2_{\varphi,0}\,\mathrm{d}y +\frac{\phi}{L} \frac{\partial}{\partial z}h_0\int_{0}^{1}c_{\varphi,0} c_{z,0}\,\mathrm{d}y= \\ &\quad\quad\quad\quad\quad\quad\quad\quad\quad\quad =  -\frac{h_0}{2}\frac{\partial p_0}{\partial \varphi} + \frac{1}{2\psi}\tau_{yx,0}|^1_0,\\
		&\phi \frac{\partial}{\partial \varphi}h_0\int_{0}^{1}c_{\varphi,0} c_{z,0}\,\mathrm{d}y +\frac{\phi^2}{L}\frac{\partial}{\partial z}h_0\int_{0}^{1}c^2_{z,0}\,\mathrm{d}y = \\
		&\quad\quad\quad\quad\quad\quad\quad\quad\quad\quad = -\frac{h_0}{2L}\frac{\partial p_0}{\partial z} + \frac{1}{2\psi}\tau_{yz,0}|^1_0.
	\end{split}
\end{equation}

By comparing the zeroth-order approximate to the initial equations, it becomes clear that only the time dependant terms within the continuity and momentum equations vanish. The same holds for the integrals, the modelling of the shear stresses and the boundary conditions. Therefore, the zeroth-order equations are equal to the non-linear partial differential equation system presented by Kuhr et al. \citep{Kuhr.Tobepublishedb}.

The first-order continuity equation yields
\begin{equation}
	\begin{split}
		&\frac{\partial}{\partial t}h_1 + \frac{\partial}{\partial \varphi}h_0\int_{0}^{1}c_{\varphi,1}\,\mathrm{d}y + \frac{\partial}{\partial \varphi}h_1\int_{0}^{1}c_{\varphi,0}\,\mathrm{d}y \;+ \\
		&\quad\quad\quad +\,\frac{\phi}{L} \frac{\partial}{\partial z}h_0\int_{0}^{1}c_{z,1}\,\mathrm{d}y + \frac{\phi}{L} \frac{\partial}{\partial z}h_1\int_{0}^{1}c_{z,0}\,\mathrm{d}y = 0.
	\end{split}
\end{equation}

The first-order approximation of the momentum equation read for circumferential direction
\begin{equation}
	\begin{split}
		&\frac{\partial}{\partial t} h_0 \int_{0}^{1} c_{\varphi,1} \, \mathrm{d}y + \frac{\partial}{\partial t} h_1 \int_{0}^{1} c_{\varphi,0} \, \mathrm{d}y \, + \\
		& + 2 \frac{\partial}{\partial \varphi} h_0 \int_{0}^{1} c_{\varphi,0} c_{\varphi,1} \, \mathrm{d}y + \frac{\partial}{\partial \varphi} h_1 \int_{0}^{1} c^2_{\varphi,0} \, \mathrm{d}y \, + \\
		& + \frac{\phi}{L} \frac{\partial}{\partial z} h_0 \int_{0}^{1} c_{\varphi,0} c_{z,1} \, \mathrm{d}y  + \frac{\phi}{L} \frac{\partial}{\partial z} h_0 \int_{0}^{1} c_{\varphi,1} c_{z,0} \, \mathrm{d}y \, +\\
		& + \frac{\phi}{L} \frac{\partial}{\partial z} h_1 \int_{0}^{1} c_{\varphi,0} c_{z,0} \, \mathrm{d}y = \\
		&\quad\quad\quad\quad\quad\quad = -\frac{h_0}{2} \frac{\partial p_1}{\partial \varphi} - \frac{h_1}{2} \frac{\partial p_0}{\partial \varphi} + \frac{1}{2\psi} \tau_{yx,1}|^1_0,
	\end{split}
\end{equation}

 and axial direction
 \begin{equation}
 	\begin{split}
 		&\phi \frac{\partial}{\partial t} h_0 \int_{0}^{1} c_{z,1} \, \mathrm{d}y \, + \phi \frac{\partial}{\partial t} h_1 \int_{0}^{1} c_{z,0} \, \mathrm{d}y \, + \\
 		& + \phi \frac{\partial}{\partial \varphi} h_0 \int_{0}^{1} c_{\varphi,0} c_{z,1} \, \mathrm{d}y + \phi \frac{\partial}{\partial \varphi} h_0 \int_{0}^{1} c_{\varphi,1} c_{z,0} \, \mathrm{d}y \, + \\
 		& + \phi \frac{\partial}{\partial \varphi} h_1 \int_{0}^{1} c_{\varphi,0} c_{z,0} \, \mathrm{d}y + 2 \frac{\phi^2}{L} \frac{\partial}{\partial z} h_0 \int_{0}^{1} c_{z,0} c_{z,1} \, \mathrm{d}y \, + \\
 		& + \frac{\phi^2}{L} \frac{\partial}{\partial z} h_1 \int_{0}^{1} c^2_{z,0} \, \mathrm{d}y \, = \\ 
 		&\quad\quad\quad\quad\quad\quad = -\frac{h_0}{2L} \frac{\partial p_1}{\partial z} - \frac{h_1}{2L} \frac{\partial p_0}{\partial z} + \frac{1}{2\psi} \tau_{yz,1}|^1_0.
 	\end{split}
 \end{equation}

The perturbation analysis is also applied to the integrals as well as the shear stresses and the boundary conditions. The perturbed integrals for the continuity and momentum equations yield
\begin{equation}
	\begin{split}
		& \int_{0}^{1}c_{\varphi,1}\,\mathrm{d}y = \frac{n_\varphi}{n_\varphi + 1} C_{\varphi,1},\\
		& \int_{0}^{1}c_{z,1}\,\mathrm{d}y = \frac{n_z}{n_z + 1} C_{z,1},\\
		& \int_{0}^{1}c_{\varphi,0}c_{\varphi,1}\,\mathrm{d}y = \frac{n_\varphi}{n_\varphi + 2} C_{\varphi,0}C_{\varphi,1} + \\
		&\quad\quad\quad\quad\quad\quad\quad\quad + \frac{n_\varphi}{2\left(n_\varphi +1\right)\left(n_\varphi+2\right)}C_{\varphi,1},\\
		& \int_{0}^{1}c_{\varphi,0}c_{z,1}\,\mathrm{d}y = \frac{n_\varphi n_z}{n_\varphi + n_z\left(n_\varphi +1\right)}C_{\varphi,0}C_{z,1} + \\
		&\quad\quad\quad\quad\quad\quad + \frac{n_z^2}{2\left(n_z + 1\right)\left[n_\varphi + n_z\left(n_\varphi + 1\right)\right]}C_{z,1},\\
		& \int_{0}^{1}c_{\varphi,1}c_{z,0}\,\mathrm{d}y = \frac{n_z n_\varphi}{n_\varphi + n_z\left(n_\varphi+1\right)} C_{z,0}C_{\varphi,1},\\
		& \int_{0}^{1}c_{z,0}c_{z,1}\,\mathrm{d}y = \frac{n_z}{n_z + 2} C_{z,0}C_{z,1}.
	\end{split}
\end{equation}

Separating the wall shear stresses $\tau_{yi,1}|^1_0$ into their corresponding directional part yields
\begin{equation}
	\tau_{yi,1}|^1_0 = \tau_{yi,\mathrm{stat},1} - \tau_{yi,\mathrm{rot},1}.
\end{equation}

By applying the binominal approximation~$(1+x)^a \approx 1 +ax$~with $|x|\ll1$, the perturbed directional wall shear stresses $\tau_{yi,\mathrm{stat},1}$ and $\tau_{yi,\mathrm{rot},1}$ read
\begin{equation} 
	\begin{split}
		& \tau_{yi,\mathrm{stat},1} = \tau_{yi,\mathrm{stat},0} \bigg[\frac{C_{\mathrm{stat},1}}{C_{\mathrm{stat},0}} + \frac{C_{i,\mathrm{stat},1}}{C_{i,\mathrm{stat},0}} + \\
		&\quad\quad\quad\quad\quad\quad\quad\quad\quad\quad+ n_f\left(\frac{h_1}{h_0} + \frac{C_{\mathrm{stat},1}}{C_{\mathrm{stat},0}}\right)\bigg],\\
		& \tau_{yi,\mathrm{rot},1} = \tau_{yi,\mathrm{rot},0} \bigg[\frac{C_{\mathrm{rot},1}}{C_{\mathrm{rot},0}} + \frac{C_{i,\mathrm{rot},1}}{C_{i,\mathrm{rot},0}} + \\
		&\quad\quad\quad\quad\quad\quad\quad\quad\quad\quad+ n_f\left(\frac{h_1}{h_0} + \frac{C_{\mathrm{rot},1}}{C_{\mathrm{rot},0}}\right)\bigg].
	\end{split}
\end{equation}

Here $C_{\mathrm{stat},1}$, $C_{\mathrm{rot},1}$ are the perturbed effective relative velocities between the stator, rotor and the fluid
\begin{equation}
	\begin{split}
		& C_{\mathrm{stat},1} = \frac{C_{\varphi,\mathrm{stat},0}C_{\varphi,\mathrm{stat},1}}{C_{\mathrm{stat},1}} + \phi^2\frac{C_{z,\mathrm{stat},0}C_{z,\mathrm{stat},1}}{C_{\mathrm{stat},1}},\\
		& C_{\mathrm{rot},1} = \frac{C_{\varphi,\mathrm{rot},0}C_{\varphi,\mathrm{rot},1}}{C_{\mathrm{rot},1}} + \phi^2\frac{C_{z,\mathrm{rot},0}C_{z,\mathrm{rot},1}}{C_{\mathrm{rot},1}},
	\end{split}
\end{equation}

and $C_{i,\mathrm{stat},1}$, $C_{i,\mathrm{rot},1}$ are the corresponding perturbed boundary layer averaged velocities
\begin{equation}
	\begin{split}
		& C_{\varphi,\mathrm{stat},1} = \frac{n_\varphi}{n_\varphi+1}C_{\varphi,1}, \quad C_{\varphi,\mathrm{rot},1} = \frac{n_\varphi}{n_\varphi+1}C_{\varphi,1},\\
		&C_{z,\mathrm{stat},1} = \frac{n_z}{n_z+1}C_{z,1},\quad C_{z,\mathrm{rot},1} = \frac{n_z}{n_z+1}C_{z,1}.\\
	\end{split}
\end{equation}

To solve the perturbed linearised partial differential equation system the perturbation analysis is applied on the boundary conditions. The perturbed pressure boundary conditions at the gap inlet and outlet read
\begin{equation}
	\begin{split}
	&p_1|_{z=0} = -2\left(1+\zeta\right)\left(\phi^2 C_{z,0}C_{z,1}\right) -2\zeta C_{\varphi,0}C_{\varphi,1},\\
	&p_1|_{z=1} = 0.
	\end{split}
\end{equation}

Furthermore, it is assumed that the perturbed pre-swirl can be neglected, i.e. $C_{\varphi,1}|_{z=0} = 0$.

As stated above, the perturbation is assumed to be harmonic in time. Thus, the perturbed  variables $C_{\varphi,1}$, $C_{z,1}$ and $p_1$ are also harmonic. This yields
\begin{equation}
	\begin{split}
		& C_{\varphi,1} = C_{\varphi,1,\mathrm{cos}} \cos\omega t + C_{\varphi,1,\mathrm{sin}}\sin\omega t,\\
		& C_{z,1}=C_{z,1,\mathrm{cos}}\cos\omega t + C_{z,1,\mathrm{sin}}\sin\omega t,\\
		& p_{1}=p_{1,\mathrm{cos}}\cos\omega t + p_{1,\mathrm{sin}}\sin\omega t.
	\end{split}
\end{equation}

The perturbed gap function is given for lateral and angular perturbed movements
\begin{equation}
	\begin{gathered}	
		h_{1} = -\left( \cos\varphi \cos\omega t + \sin\varphi \sin\omega t \right),\\
		h_{1} = -\left( z - z_T \right) \frac{L}{\psi} \left( \cos\varphi \cos\omega t - \sin\varphi \sin\omega t \right).
	\end{gathered}
\end{equation}

Assuming a harmonic perturbation, the time dependant terms within the continuity and momentum equation vanish and the first-order linearised partial differential equation system can be solved. To solve the equation system the same SIMPLE-C algorithm as for solving the zeroth-order equation system is used, cf. \citep{Kuhr.Tobepublishedb}. By integrating the first-order pressure distribution the corresponding forces and torques acting on the rotor are determined by
\begin{equation}
	\begin{split}	
		& F_{X,1}=-\int_{0}^{1}\int_{0}^{2\pi}p_1\cos\varphi\,\mathrm{d}\varphi\,\mathrm{d}z,\\
		& F_{Y,1}=-\int_{0}^{1}\int_{0}^{2\pi}p_1\sin\varphi\,\mathrm{d}\varphi\,\mathrm{d}z,\\
		& M_{X,1}=\int_{0}^{1}\int_{0}^{2\pi}p_1\left(z-z_T\right)\sin\varphi\,\mathrm{d}\varphi\,\mathrm{d}z,\\
		& M_{Y,1}=-\int_{0}^{1}\int_{0}^{2\pi}p_1\left(z-z_T\right)\cos\varphi\,\mathrm{d}\varphi\,\mathrm{d}z.\\
	\end{split}
\end{equation}

The rotordynamic coefficients are obtained by calculating the induced forces and torques at different percessional frequencies $\omega$ and performing a least mean square identification procedure. Here, the lateral and angular movements are treated separately. As an example, the force $F_X$ and torque $M_X$ generated by lateral displacements are
\begin{equation}
	\begin{split}	
		& -F_{X}= K_{XX}X+K_{XY}Y+\\
		&\quad\quad\quad\quad\quad +C_{XX}\dot{X}+C_{XY}\dot{Y}+M_{XX}\ddot{X}+M_{XY}\ddot{Y},\\
		& -M_{X}= K_{\alpha X}X+K_{\alpha Y}Y+\\
		&\quad\quad\quad\quad\quad +C_{\alpha X}\dot{X}+C_{\alpha Y}\dot{Y}+M_{\alpha X}\ddot{X}+M_{\alpha Y}\ddot{Y}.
	\end{split}
\end{equation}

Evaluating the perturbed forces and torques at $\omega t=0$ and $\omega t=\pi/2$ the $\cos$, $\sin$ terms vanish and the coefficients can be extracted.
\begin{equation}
	\begin{split}
		&-F_{X,1}\left(\omega t = 0\right) = K_{XX}+C_{XY}\omega - M_{XX}\omega^2,\\
		&-F_{X,1}\left(\omega t = \pi/2\right) = K_{XY} - C_{XX}\omega - M_{XY}\omega^2,\\
		&-M_{X,1}\left(\omega t = 0\right) = K_{\alpha X}+C_{\alpha Y}\omega - M_{\alpha X}\omega^2,\\
		&-M_{X,1}\left(\omega t = \pi/2\right) = K_{\alpha Y} - C_{\alpha X}\omega - M_{\alpha Y}\omega^2.
	\end{split}
\end{equation}

\section{Validation}
For the purpose of validation, the rotordynamic coefficients for translational displacements are compared to the numerical results published by Nordmann \& Dietzen \citep{Nordmann.1988}, Nelson \& Nguyen \citep{Nelson.1988b} and Simon \& Fr\^{e}ne \citep{Simon.1992}. Here, the influence of eccentricity on the rotordynamic coefficients is investigated for an annulus with length $L = 0.5$, a modified Reynolds number $Re_\varphi^*=0.043$, a pressure difference $\Delta p := 2\Delta \tilde{p}/\left(\tilde{\varrho}\tilde{\Omega}^2\tilde{R}^2\right) = 1.78$ and a pre-swirl $C_\varphi|_{z=0}=0.3$. Nordmann \& Dietzen \citep{Nordmann.1988} use a three- dimensional finite-difference method, solving the Navier-Stokes and continuity equation in combination with a $k$-$\epsilon$ turbulence model, whereas Nelson \& Nguyen \citep{Nelson.1988b} present a method using fast Fourier transforms to integrate the governing equation system resulting in hydrodynamic forces and rotordynamic coefficients. The results by Simon \& Fr\^{e}ne \citep{Simon.1992} are generated by a similar approach to the one presented here. The main difference is the treatment of the integrals of the partial differential equation system. While the here presented method uses ansatz functions to describe the velocity profile before integration, Simon \& Fr\^{e}ne \citep{Simon.1992} uses parabolic functions for the integral itself.

For angular displacements, the model is compared to data by Childs \citep{Childs.1982} and San And\'{e}s \citep{SanAndres.1993}. The authors investigate the influence of gap length on the rotordynamic torque coefficients due to angular displacement for a concentric annulus with the modified Reynolds number $Re_\varphi^* = 0.029$, a pressure difference $\Delta p = 8.38$ and no pre-swirl $C_\varphi|_{z=0}=0$. Childs \citep{Childs.1982} uses a bulk-flow approach for centred finite-length seals, whereas San And\'{e}s \citep{SanAndres.1993} presents a two-dimensional calculation method considering fully developed flow.

\begin{figure*}
	\centering
	\includegraphics[scale=0.87]{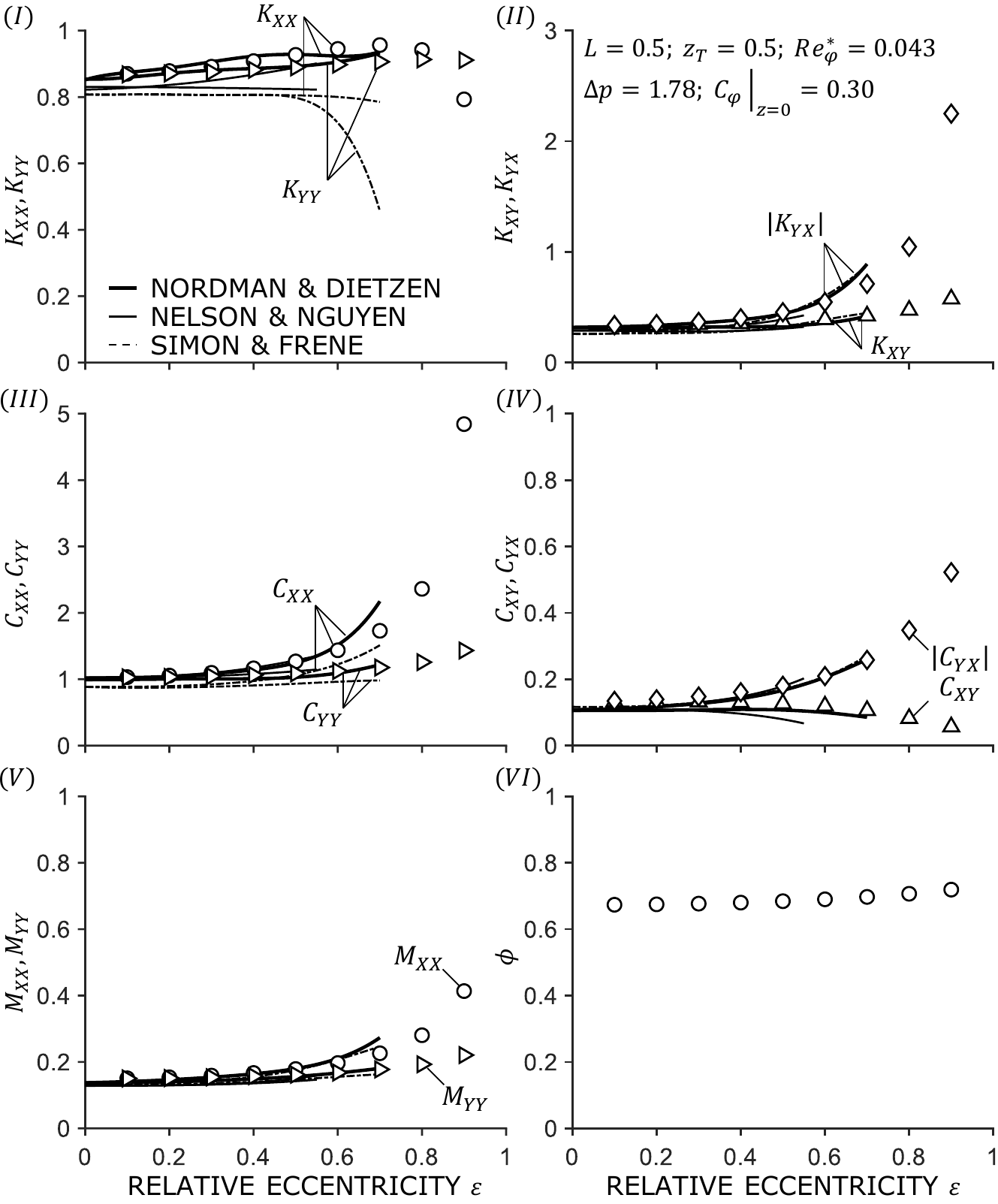}
	\caption{Rotordynamic coefficients determined by the CAPM for translational displacement compared to the numeric results by Nordmann \& Dietzen \citep{Nordmann.1988}, Nelson \& Nguyen \citep{Nelson.1988b} and Simon \& Fr\^{e}ne.}
	\label{fig:figure_validation_cylindrical}
\end{figure*}

Figure \ref{fig:figure_validation_cylindrical} shows the comparison of the influence of eccentricity on the rotordynamic coefficients for translational displacement. Here, the lines are the numeric results by Nordmann \& Dietzen \citep{Nordmann.1988}, Nelson \& Nguyen \citep{Nelson.1988b} and Simon \& Fr\^{e}ne , whereas the markers represent the Clearance-Averaged Pressure Model. The figure compares (I) the direct stiffness $K_{XX}$, $K_{YY}$, (II) the cross-coupled stiffness $K_{XY}$, $K_{YX}$, (III) the direct damping $C_{XX}$, $C_{YY}$, (IV) the cross-coupled damping $C_{XY}$, $C_{YX}$ and (V) the direct inertia $M_{XX}$, $M_{YY}$ coefficients. In addition, the influence of eccentricity on (VI) the flow number is given. 

It exhibits a good agreement of the CAPM with the data obtained from the literature. Major differences in the calculation methods are only apparent for the direct stiffness $K_{XX}$,  $K_{YY}$. Here, the results presented by Nordman \& Dietzen \citep{Nordmann.1988} and Simon \& Fr\^{e}ne \citep{Simon.1992} are in good agreement with the data obtained by the presented method. However, the results of Nelson \& Nguyen \citep{Nelson.1988b} differ at higher eccentricities. The predicted decrease of the curves starts at much lower eccentricities and is far more severe than the results predicted  by the other authors as well the CAPM. In addition, the eccentricity influence on the flow number is given. The flow number increases with increasing eccentricity. This is due to the altered friction losses within the annulus at eccentric operation conditions and a constant axial pressure difference $\Delta p$. 

\begin{figure*}
	\centering
	\includegraphics[scale=0.87]{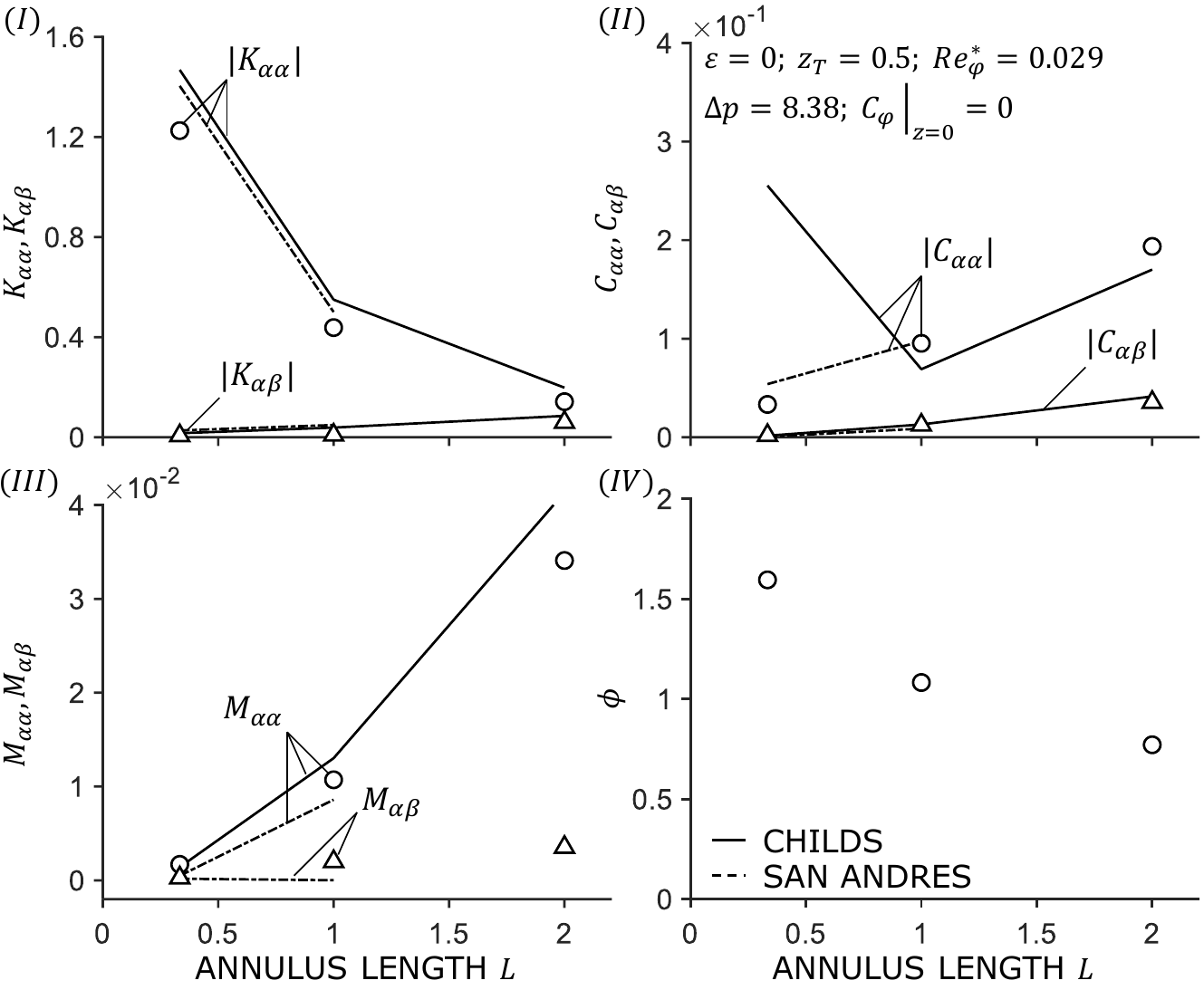}
	\caption{Rotordynamic coefficients determined by the CAPM for angular displacement compared to the numeric results by Childs \citep{Childs.1982} and San And\'{e}s \citep{SanAndres.1993}.}
	\label{fig:figure_validation_conical}
\end{figure*}

Figure \ref{fig:figure_validation_conical} shows the comparison of the influence of the annulus length on the rotordynamic coefficients for angular displacement. The figure compares (I) the direct and cross-coupled stiffness $|K_{\alpha \alpha}|$, $K_{\alpha \beta}$, (II) the direct and cross-coupled damping $C_{\alpha\alpha}$, $C_{\alpha\beta}$ and (III) the direct and cross-coupled inertia $M_{\alpha\alpha}$, |$M_{\alpha\beta}|$. In addition, the influence of annulus length on (IV) the flow number is given. 

Again it exhibits a good agreement of the CAPM with the data obtained from the literature. Minor differences in the calculation methods are only apparent for the direct damping $C_{\alpha\alpha}$. Here, the results presented by San Andr\'{e}s \citep{SanAndres.1993} are in good agreement with the data obtained by the presented method. However, the results of Childs \citep{Childs.1982} differ at an annulus length $L=0.33$. The value is much higher than predicted by San Andr\'{e}s \citep{SanAndres.1993} and the presented method CAPM. Furthermore, the predicted value seems not to follow the overall trend, i.e. decreased direct damping $C_{\alpha\alpha}$ at decreasing gap length. Compared to the results presented here and the results obtained by San Andr\'{e}s \citep{SanAndres.1993},  the calculation seems to be somewhat inconsistent. In addition, the length influence on the flow number is given. The flow number decreased with increasing length. This is due to the increased friction losses within the annulus at concentric operation conditions and constant axial pressure difference $\Delta p$. 

\section{Parameter study}
Considering the good agreement of the presented model with the data from the literature, an extensive parameter study is carried out, focusing on the influence of the annulus length, the eccentricity, the centre of rotation, the modified Reynolds number, the flow number and the pre-swirl on the force and the torque characteristics. In the following, only the results for the influence of the annulus length, the modified Reynolds number and the flow number are given. It is shown that the annulus length, the modified Reynolds number and the flow number are crucial when determining the relevance of the hydraulic tilt and torque coefficients. The remaining results, the influence of eccentricity, the centre of rotation and the pre-swirl are given in the appendix \ref{sec:appendix}.

\subsection{Influence of the annulus length}
\begin{figure*}
	\centering
	\includegraphics[scale=0.87]{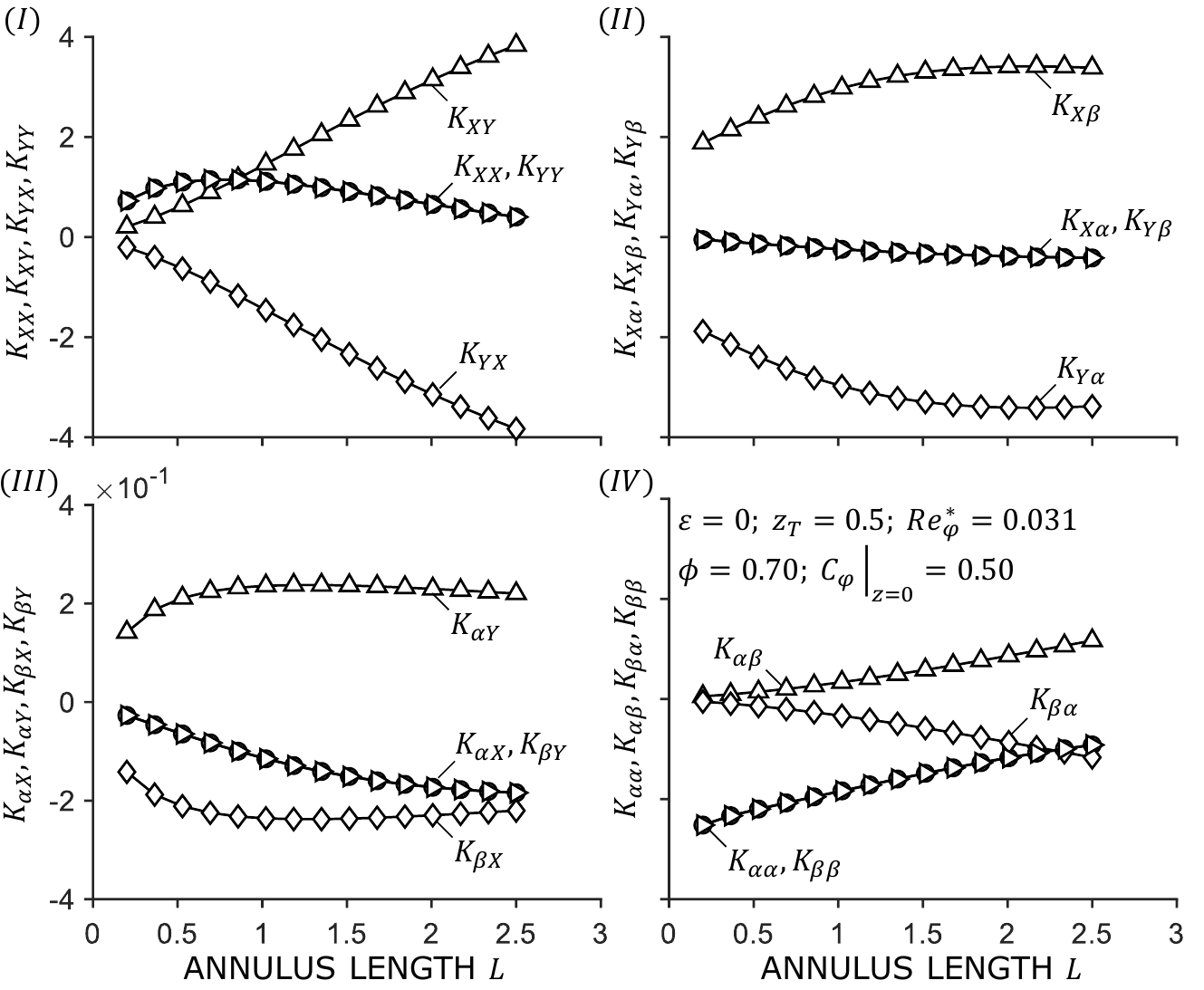}
	\caption{Influence of annulus length on the stiffness due to translational and angular excitation. (I) Stiffness due to translational excitation by the hydraulic forces. (II) Stiffness due to angular excitation by the hydraulic forces. (III) Stiffness due to translational excitation by the hydraulic torques. (IV) Stiffness due to angular excitation by the hydraulic torques.}
	\label{fig:figure_results_length_stiffness_all}
\end{figure*}

\begin{figure}
	\centering
	\includegraphics[scale=0.87]{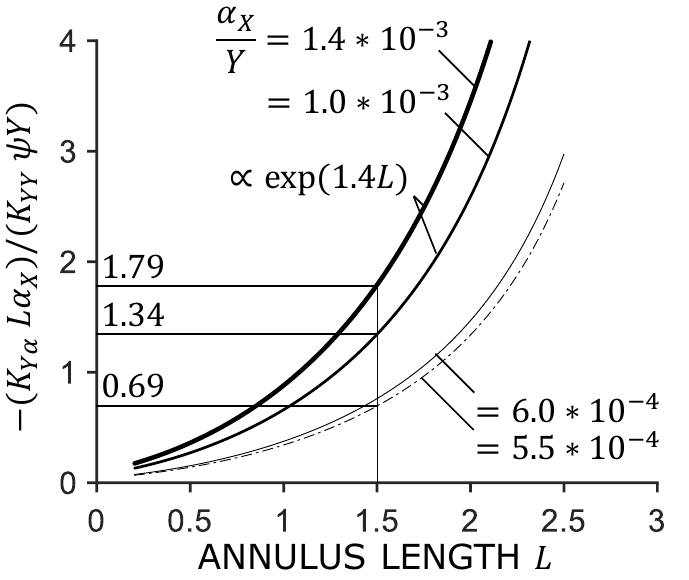}
	\caption{Influence of the annulus length and the ratio of angular to translational excitation $\alpha_X/Y$ on the ratio of tilt to translational stiffness coefficients.}
	\label{fig:figure_results_length_influenceTilt}
\end{figure}

\begin{figure*}
	\centering
	\includegraphics[scale=0.87]{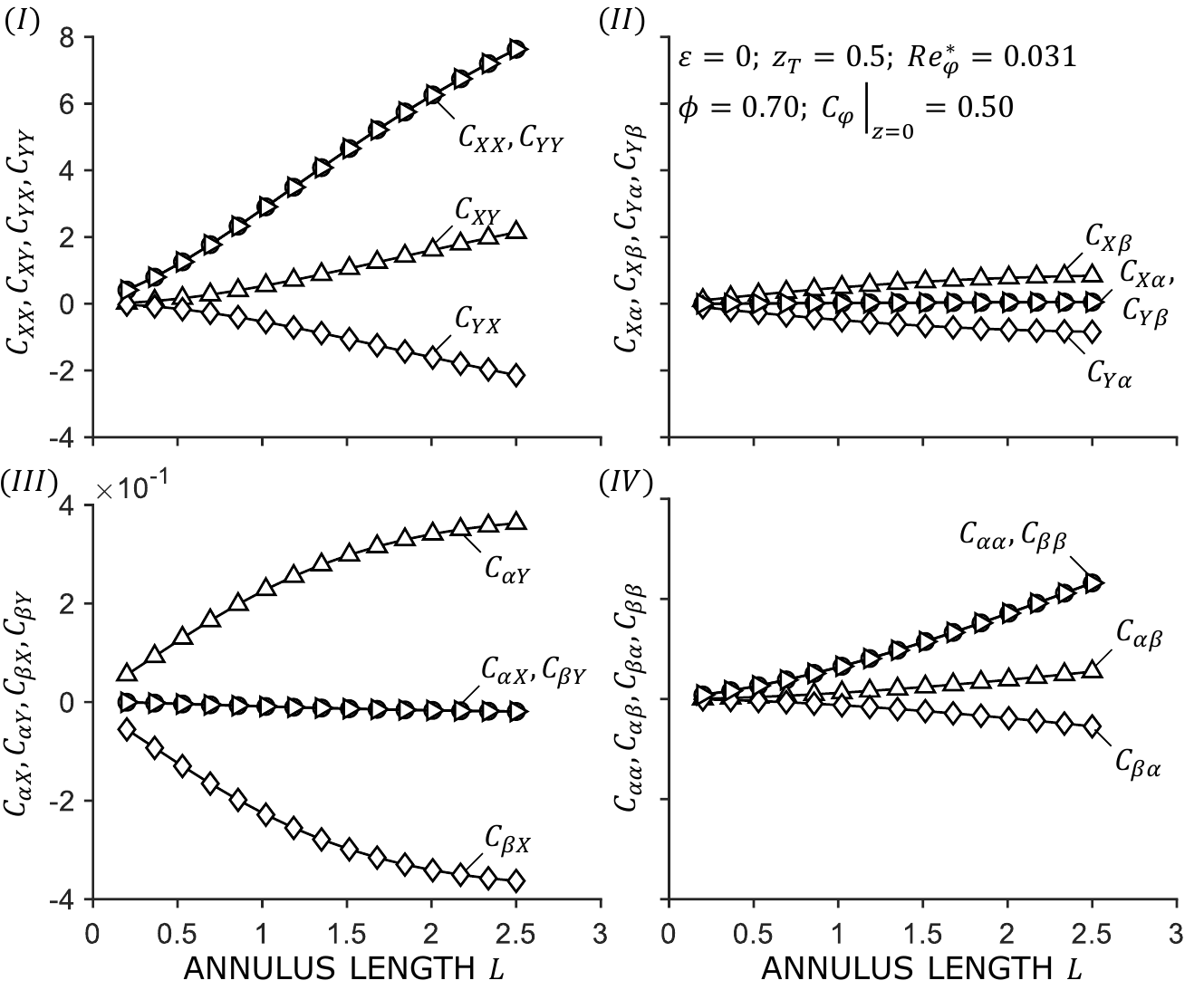}
	\caption{Influence of annulus length on the damping due to translational and angular excitation. (I) Damping due to translational excitation by the hydraulic forces. (II) Damping due to angular excitation by the hydraulic forces. (III) Damping due to translational excitation by the hydraulic torques. (IV) Damping due to angular excitation by the hydraulic torques.}
	\label{fig:figure_results_length_damping_all}
\end{figure*}

\begin{figure*}
	\centering
	\includegraphics[scale=0.87]{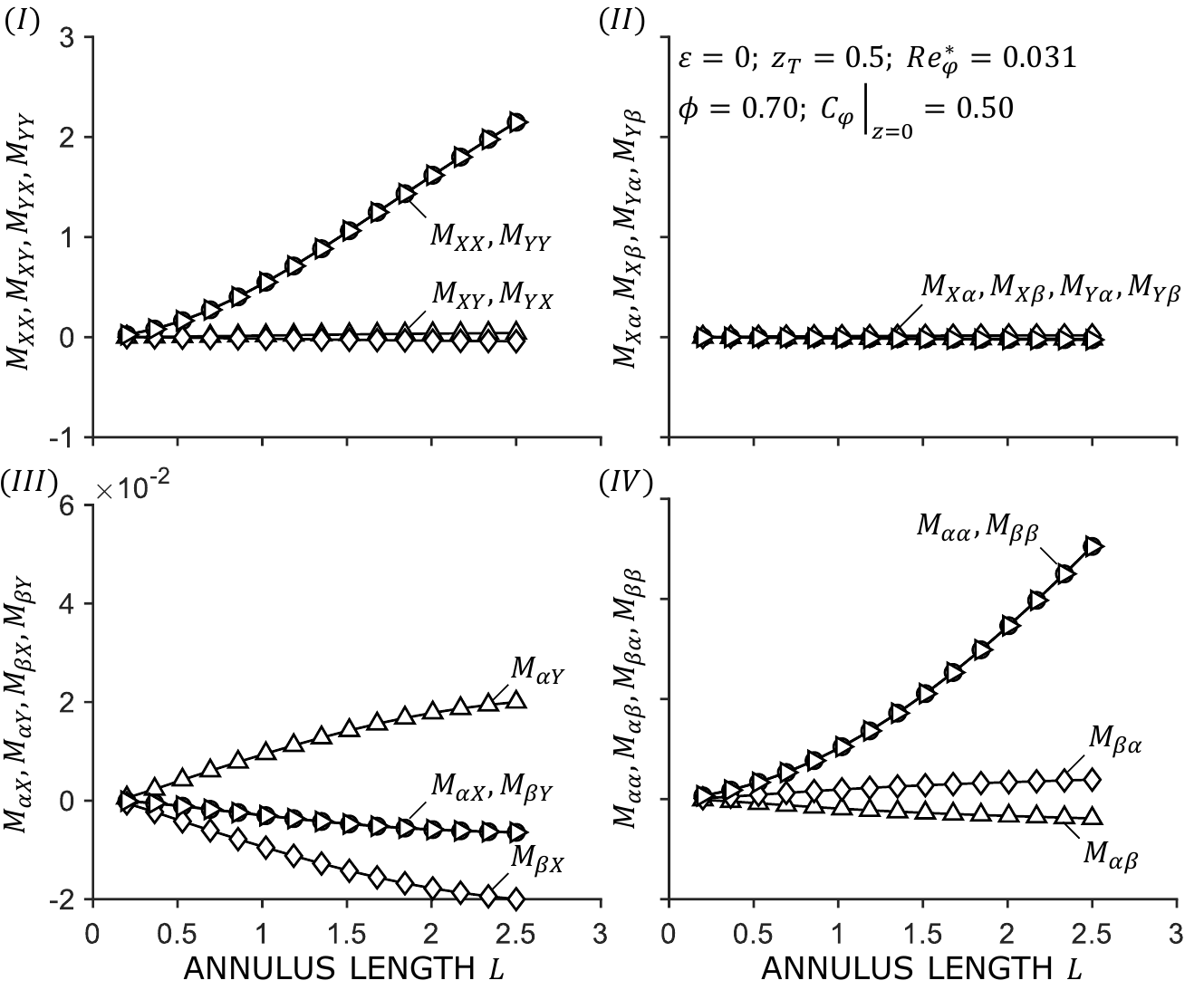}
	\caption{Influence of annulus length on the inertia due to translational and angular excitation. (I) Inertia due to translational excitation by the hydraulic forces. (II) Inertia due to angular excitation by the hydraulic forces. (III) Inertia due to translational excitation by the hydraulic torques. (IV) Inertia due to angular excitation by the hydraulic torques.}
	\label{fig:figure_results_length_inertia_all}
\end{figure*}

In the following, the influence of the annulus length is investigated. Figures \ref{fig:figure_results_length_stiffness_all} to \ref{fig:figure_results_length_inertia_all} show the rotordynamic force and torque coefficients for translational and angular excitation. The examined annulus is operated at concentric conditions, i.e. $\varepsilon = 0$ at a modified Reynolds number $Re_\varphi = 0.031$ and flow number $\phi = 0.7$. The pre-swirl is $C_\varphi|_{z=0}=0.5$, whereas the fulcrum lies in the centre of the annular gap, i.e. $z_T = 0.5$. All four submatrices are skew-symmetric for the chosen concentric operation point, i.e $K_{I..IV}^T, C_{I..IV}^T, M_{I..IV}^T = -K_{I..IV}, -C_{I..IV}, -M_{I..IV}$.\\ Therefore, it is sufficient to focus on one of the direct and one of the cross-coupled stiffness coefficients for each matrix. However, for the sake of completeness all coefficients are shown in the figures.\\

Figure \ref{fig:figure_results_length_stiffness_all} shows the influence of the annulus length on the direct and cross-coupled stiffness due to translational excitation by the hydraulic forces (I), the direct and cross-coupled stiffness due to angular excitation caused by the hydraulic forces (II), the direct and cross-coupled stiffness due to translational excitation by the hydraulic torques (III) and the direct and cross-coupled stiffness due to angular excitation by the hydraulic torques (IV). Focusing on submatrix (I) the cross-coupled stiffness $K_{XY}$, $|K_{YX}|$ increase linearly with increasing annulus length, whereas the direct stiffness coefficients $K_{XX}$, $K_{YY}$ initially increase, reaching a maximum at an annulus length of $L \approx 0.6$. By further increasing the annulus length, the direct stiffness decreases. It should be noted that as the gap becomes longer, the cross-coupled stiffness becomes larger than the direct stiffness. This is particularly interesting, as it can be crucial to stability analysis. In contrast to the the direct and cross-coupled stiffness of submatrix (I), the direct and cross-coupled stiffness of submatrix (II) show an asymptotic behaviour when increasing the annulus length. Fist, focusing on the direct stiffness, i.e $K_{X\alpha}$, $K_{Y\beta}$, the stiffness decreases with increasing gap length. The curve continues to flatten reaching a value of $K_{X\alpha}$, $K_{Y\beta} = -0.41$ at $L = 2.5$. Second, examining the cross-coupled coefficients $K_{X\beta}$, $|K_{Y\alpha}|$, the stiffness increases with increasing annulus length up to an annulus length of $L=1.5$. By further increasing the length, the curves show an asymptotic behaviour for a value $K_{X\beta}$, $|K_{Y\alpha}| = 3.40$. It should be noted that the cross-coupled coefficients of submatrix (II) are in the same order of magnitude as the cross- coupled stiffness of submatrix (I). This is of particular interest when examining the relevance of the hydraulic tilt and torque coefficients, cf. figure \ref{fig:figure_results_length_influenceTilt}. The direct and cross-coupled stiffness of submatrix (III) behave similarly to the ones of submatrix (II). The direct stiffness $K_{\alpha X}$, $K_{\beta Y}$, being one order of magnitude smaller than direct stiffness $K_{X\alpha}$, $K_{Y\beta}$, decrease with increasing annulus length, whereas the cross-coupled stiffness $K_{\alpha Y}$, $|K_{\beta X}|$ first increase with increasing length up to $L=0.75$. For annulus lengths longer than $L>0.75$, the cross-coupled stiffness stagnates and slightly decreases. Focusing on the fourth submatrix (IV) the direct $K_{\alpha\alpha}$, $K_{\beta\beta}$ as well as the cross-coupled stiffness $|K_{\beta\alpha}|$, $K_{\alpha\beta}$ increase when increasing the annulus length. \\

As mentioned before, the relevance of the tilt and torque coefficients can be determined by focusing on the cross-coupled stiffness of submatrix (II). \cite{Childs.1993} states that the additional coefficients become relevant at an annulus length greater than $L=1.5$. When determining the relevance of tilt and torque coefficients the focus lies on the $Y$ component of the induced forces while translational motion in $X$-direction as well as angular motion around the $Y$-axis is prohibited. This yields
\begin{equation}
    -F_Y = K_{YY}Y + K_{Y\alpha}\alpha.
\end{equation}

By defining an effective stiffness
\begin{equation}\label{eqn:results_effective_stiffness}
    \begin{split}
        K_{\mathrm{eff}}:=K_{YY}&\left( 1 + \frac{K_{Y\alpha}\alpha}{K_{YY}Y}\right) =  \\&= K_{YY}\left( 1 + \frac{K_{Y\alpha}}{K_{YY}}\frac{L}{\psi}\frac{\alpha_X}{Y}\right) 
    \end{split}
\end{equation}

 the relevance of the additional coefficients can be studied. If the quotient of $K_{Y\alpha}\alpha/\left(K_{YY}Y\right)$ is small, only the forces due to translational motion are relevant as it is for small annuli. By examining equation \ref{eqn:results_effective_stiffness} it becomes clear that an overall threshold for the relevance of the additional rotordynamic coefficients is a strong simplification. Instead, the quotient is inversely dependant on the slenderness of the gap, i.e. $L/\psi$, and the ratio of the angular and translational excitation $\alpha_X / Y$. Childs uses a constant ratio of $\alpha_X/Y = 5.5*10^{-4}$ to calculate the overall length threshold of $L/R=1.5$. This corresponds to an excitation angle of $\alpha_X = 3.10*10^{-3}\,\mathrm{degree}$ and a translational excitation of $36\, \mathrm{\mu m}$, cf. \cite{Childs.1993}. Furthermore, any additional influence besides the annulus length is neglected.\\
 
Figure \ref{fig:figure_results_length_influenceTilt} shows the ratio of tilt to translational stiffness coefficients, i.e. $|K_{Y\alpha}L\,\alpha_X/\left(K_{YY} \psi Y\right)|$ versus the annulus length for different $\alpha_X/Y$. It exhibits increasing influence of the additional rotordynamic coefficients with increasing annulus length. Here, the relevance of the additional tilt and torque coefficients increase proportionally $\propto \exp{(1.4L)}$. Furthermore, it shows a strong influence on the ratio of angular to translational excitation. Focusing on the ratio chosen by Childs, the cross-coupled stiffness of submatrix (II) $K_{Y\alpha}$ is $0.69$ of the the direct stiffness $K_{YY}$ at an annulus length of $L=1.5$. This means that the stiffness due to the tilt accounts for approximately $41\,\%$ of the total stiffness. By increasing the ratio of angular to translational excitation to $\alpha_X/Y = 1.0*10^{-3}$ the cross-coupled stiffness of submatrix (II) $K_{Y\alpha}$ is $1.34$ times as great as the direct stiffness $K_{YY}$, increasing the contribution of the tilt coefficients to the overall stiffness to $57\,\%$. It should be noted that this corresponds to an increased excitation angle of $\alpha_X = 6.0*10^{-3}\,\mathrm{degree}$. Therefore, an overall threshold of $L/R = 1.5$ to describe the relevance of the additional rotordynamic coefficients is insufficient. Rather, the operating conditions of the turbomachinery need to be taken into account.\\

Figure \ref{fig:figure_results_length_damping_all} shows the influence of the annulus length on the direct and cross-coupled damping due to translational excitation by the hydraulic forces (I), the direct and cross-coupled damping due to angular excitation by the hydraulic forces (II), the direct and cross-coupled damping due to translational excitation by the hydraulic torques (III) and the direct and cross-coupled damping due to angular excitation by the hydraulic torques (IV). It exhibits that the direct and cross-coupled damping due to translational excitation by the hydraulic forces are one order of magnitude higher than the other damping coefficients in the sub-matrices (II) to (IV).
First, focusing on the damping coefficients of submatrix (I) both, the direct $C_{XX}$, $C_{YY}$ and cross-coupled damping $C_{XY}$, $|C_{YX}|$ linearly depend on the annulus length. Here, the direct coefficients increase faster than the cross-coupled coefficients. It is noted that the direct damping coefficients are of particular interest when evaluating the stability of the flow inside the annulus. Second, focusing on the direct and cross-coupled damping coefficients of submatrix (II), the direct damping coefficients $C_{X\alpha}$, $C_{Y\beta}$ are not influenced by an increasing annulus length, whereas the cross-coupled damping coefficients $C_{X\beta}$, $|C_{Y\alpha}|$ slightly increase with increasing length. The damping coefficients of submatrix (III) are in the same order of magnitude as the coefficients of submatrix (II). Similar to the direct damping $C_{X_\alpha}$, $C_{Y\beta}$, the direct damping coefficients $C_{\alpha X}$, $C_{\beta Y}$ are almost independent of the annulus length. In contrast, the cross-coupled damping coefficients $C_{\alpha Y}$, $|C_{\beta X}|$ increase with increasing gap length. Finally, the direct and cross-coupled damping coefficients of submatrix (IV) are investigated. The direct damping coefficients $C_{\alpha\alpha}$, $C_{\beta\beta}$ as well as the cross-coupled damping coefficients $C_{\alpha\beta}$, $|C_{\beta\alpha}|$ increase with increasing annulus length. Here, the direct coefficients are greater in value than the cross-coupled coefficients.\\

Figure \ref{fig:figure_results_length_inertia_all} shows the influence of the annulus length on the direct and cross-coupled inertia due to translational excitation by the hydraulic forces (I), the direct and cross-coupled inertia due to angular excitation by the hydraulic forces (II), the direct and cross-coupled inertia due to translational excitation by the hydraulic torques (III) and the direct and cross-coupled inertia due to angular excitation by the hydraulic torques (IV).
First, focusing on the inertia coefficients of submatrix (I), both, the direct and cross-coupled inertia coefficients linearly increase with the annulus length. Here, the cross-coupled inertia coefficients are two orders of magnitude smaller than the direct inertia coefficients. As a result, most of the literature neglects the cross-coupled inertia coefficients compared to the direct coefficients. Second, focusing on the direct and cross-coupled inertia coefficients of submatrix (II), the coefficients are in the same order of magnitude as the cross-coupled inertia coefficients of submatrix (I). Therefore, no significant trend can be observed when increasing the annulus length. The inertia coefficients of submatrix (III) are in the same order of magnitude as the coefficients of submatrix (II). The direct inertia $M_{\alpha X}$, $M_{\beta Y}$ decreases slightly with increasing annulus length, whereas the 
cross-coupled inertia $\MAY$, $|\MBX|$ increases with the length. Finally, the direct and cross-coupled inertia coefficients of submatrix (IV) are investigated. The direct inertia coefficients $M_{\alpha\alpha}$, $M_{\beta\beta}$ as well as the the cross-coupled coefficients $|M_{\alpha \beta}|$, $M_{\beta \alpha}$ increase with increasing annulus length. Here, the direct coefficients are one order of magnitude greater than the cross-coupled coefficients.\\

In summary, the following statements can be made:
\begin{itemize}
    \item The cross-coupled stiffness, the direct damping and the direct inertia of submatrix (I) increase linearly with the annulus length.
    \item The cross-coupled stiffness of submatrix (II) shows an asymptotic behaviour.
    \item The relevance of the hydraulic tilt and torque coefficients increases proportionally $\propto \exp{(1.4L)}$.
\end{itemize}

\subsection{Influence of the modified Reynolds number}
\begin{figure*}
	\centering
	\includegraphics[scale=0.87]{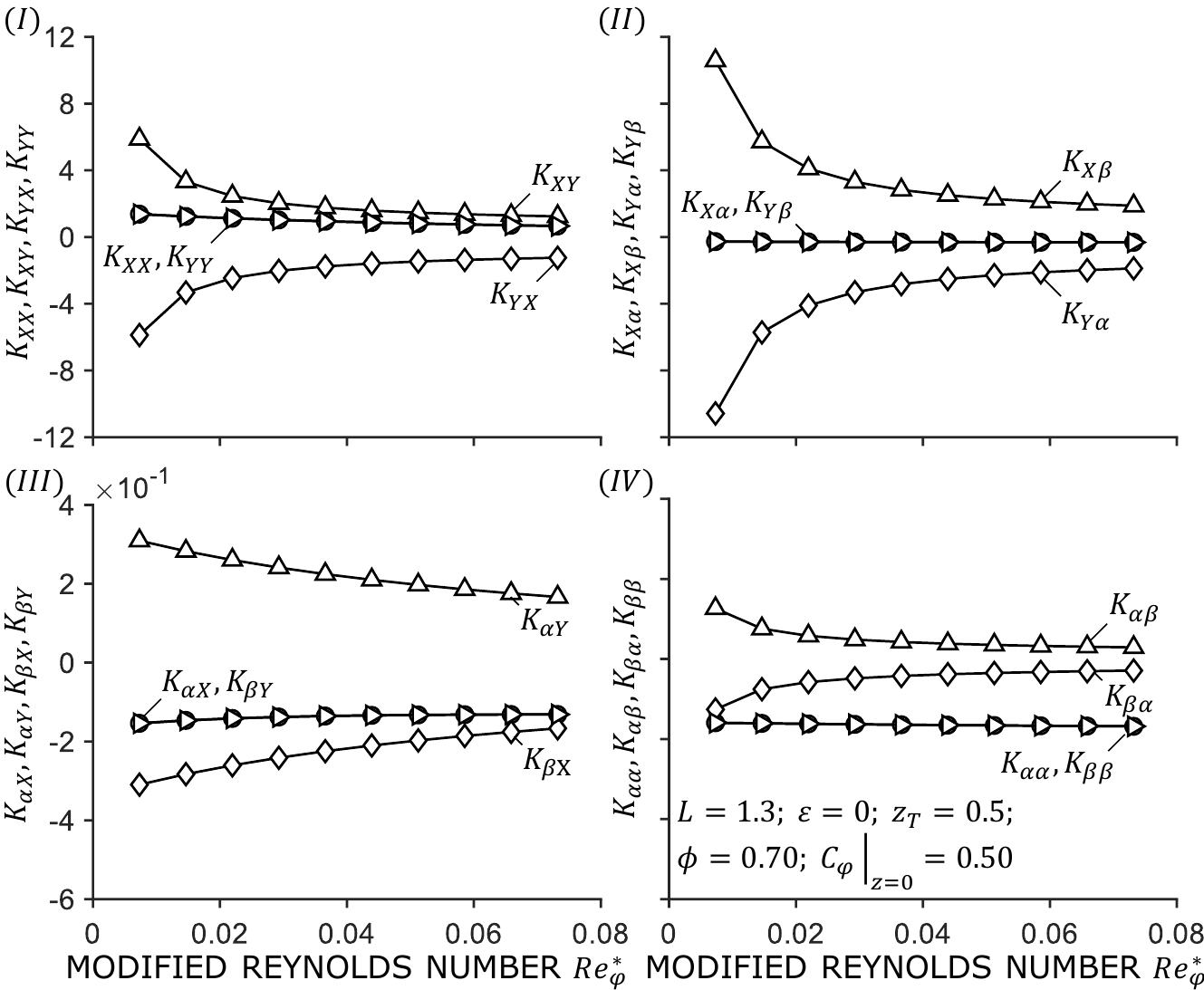}
	\caption{Influence of the modified Reynolds number on the stiffness due to translational and angular excitation. (I) Stiffness due to translational excitation by the hydraulic forces. (II) Stiffness due to angular excitation by the hydraulic forces. (III) Stiffness due to translational excitation by the hydraulic torques. (IV) Stiffness due to angular excitation by the hydraulic torques.}
	\label{fig:figure_results_psiRePhi_stiffness_all}
\end{figure*}

\begin{figure}
	\centering
	\includegraphics[scale=0.87]{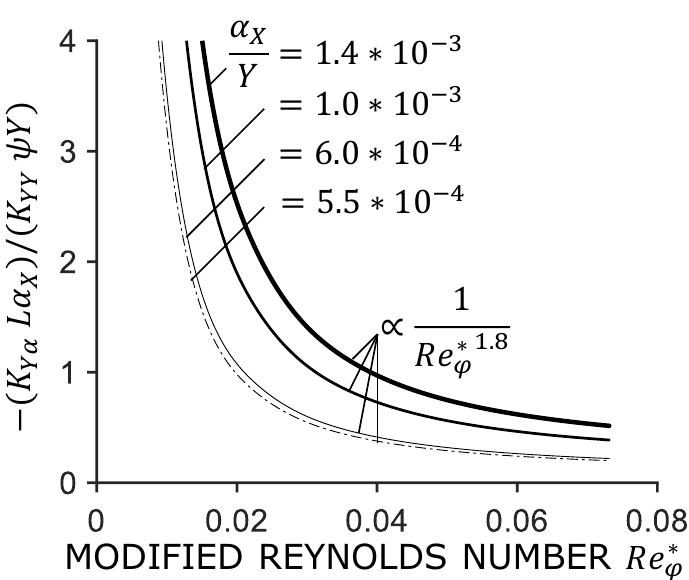}
	\caption{Influence of the modified Reynolds number and the ratio of angular to translational excitation $\alpha_X/Y$ on the ratio of tilt to translational stiffness coefficients.}
	\label{fig:figure_results_psiRePhi_influenceTilt}
\end{figure}

\begin{figure*}
	\centering
	\includegraphics[scale=0.87]{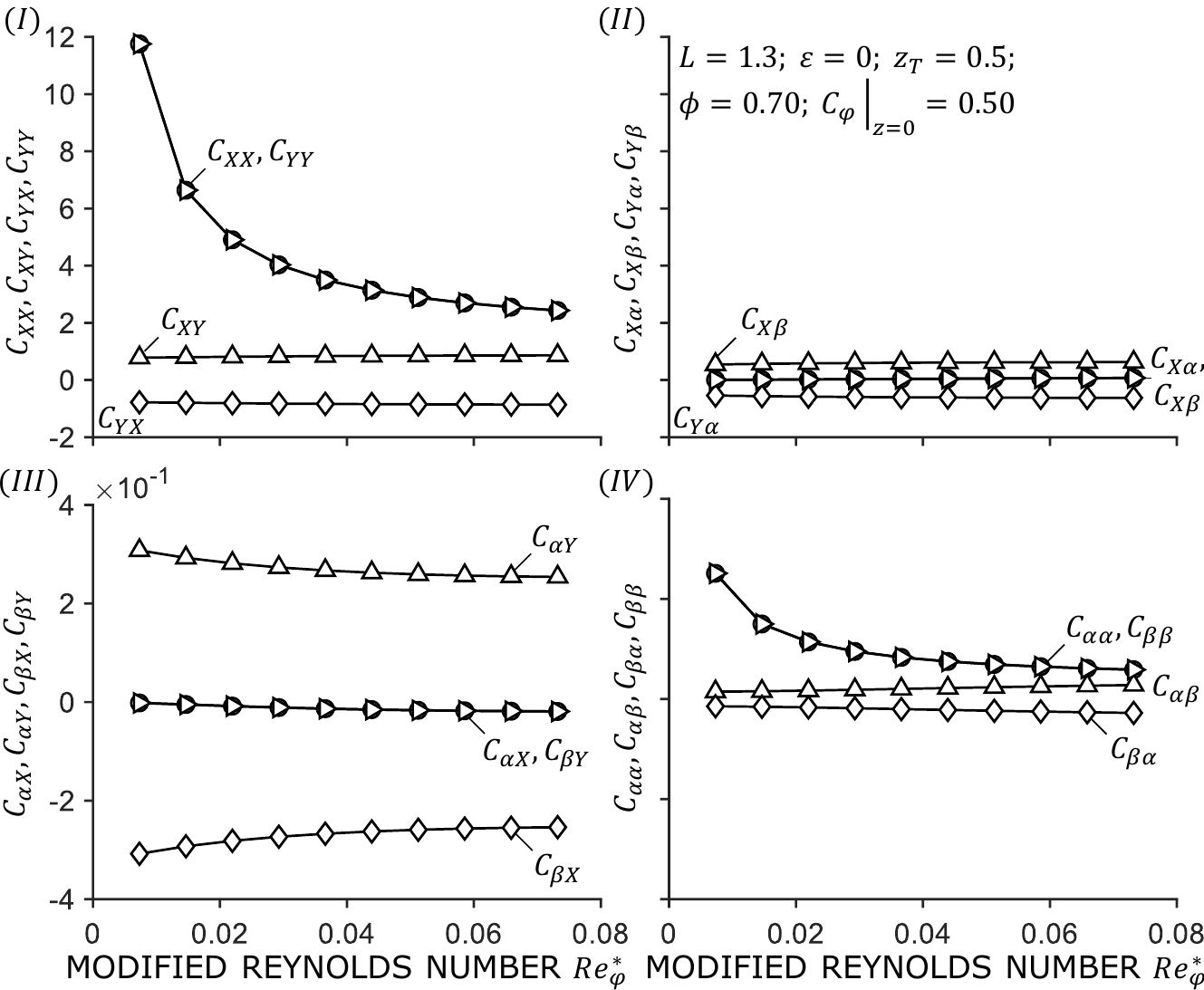}
	\caption{Influence of the modified Reynolds number on the damping due to translational and angular excitation. (I) Damping due to translational excitation by the hydraulic forces. (II) Damping due to angular excitation by the hydraulic forces. (III) Damping due to translational excitation by the hydraulic torques. (IV) Damping due to angular excitation by the hydraulic torques.}
	\label{fig:figure_results_psiRePhi_damping_all}
\end{figure*}

\begin{figure*}
	\centering
	\includegraphics[scale=0.87]{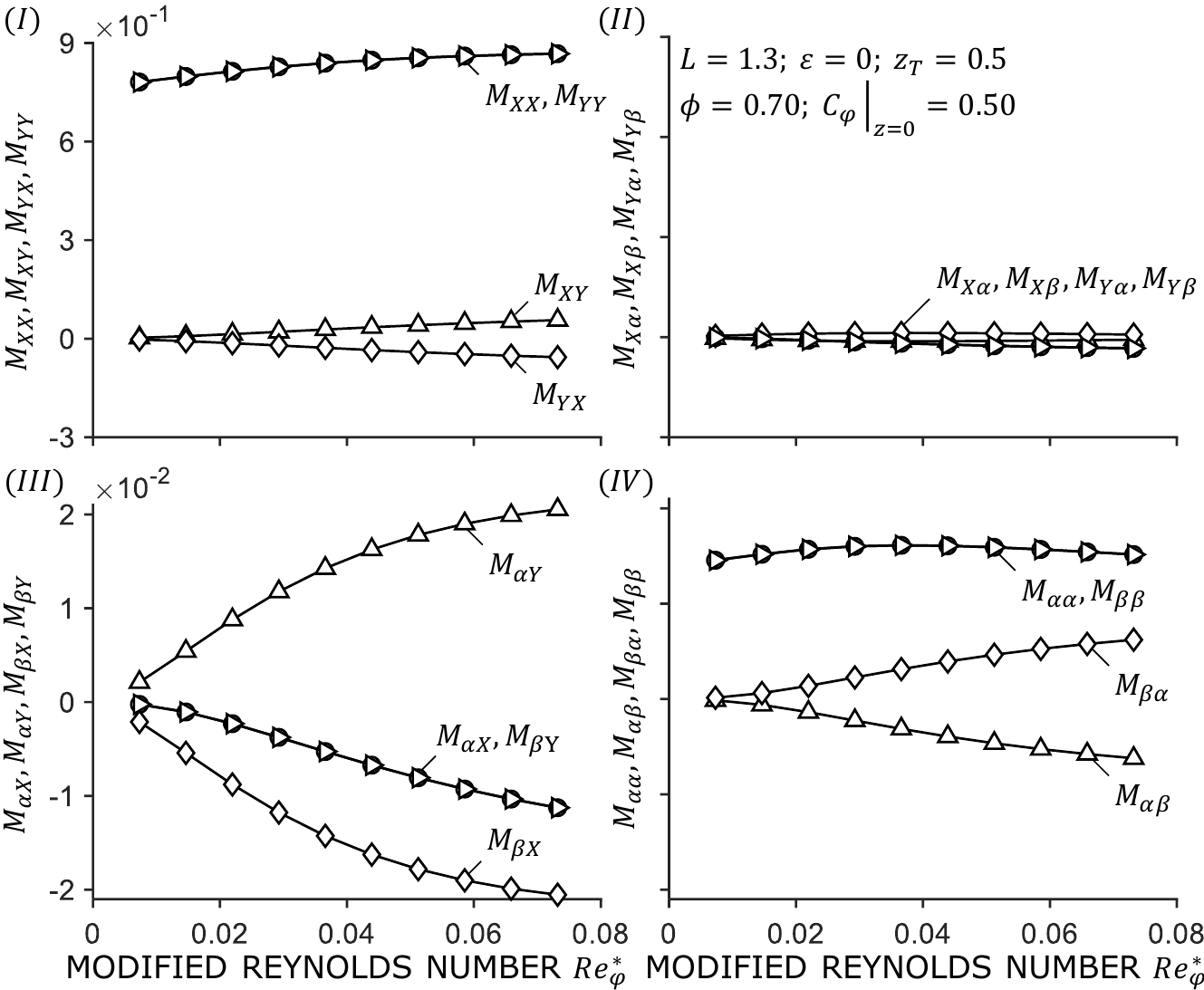}
	\caption{Influence of the modified Reynolds number on the inertia due to translational and angular excitation. (I) Inertia due to translational excitation by the hydraulic forces. (II) Inertia due to angular excitation by the hydraulic forces. (III) Inertia due to translational excitation by the hydraulic torques. (IV) Inertia due to angular excitation by the hydraulic torques.}
	\label{fig:figure_results_psiRePhi_inertia_all}
\end{figure*}

In the following the influence of the modified Reynolds number on the rotordynamic coefficients is investigated. Figures \ref{fig:figure_results_psiRePhi_stiffness_all} to \ref{fig:figure_results_psiRePhi_inertia_all} give the force and torque coefficients for translational and angular excitation. The annulus of length $L = 1.3$ is operated at concentric conditions, i.e. $\varepsilon = 0$ with a flow number $\phi = 0.7$. The pre-swirl before the annulus is set to $C_\varphi|_{z=0}=0.5$ and the fulcrum lies in the centre of the annular gap, i.e. $z_T = 0.5$.\\

Figure \ref{fig:figure_results_psiRePhi_stiffness_all} shows the influence of the modified Reynolds number on the direct and cross-coupled stiffness due to translational excitation by the hydraulic forces (I), the direct and cross-coupled stiffness due to angular excitation by the hydraulic forces (II), the direct and cross-coupled stiffness due to translational excitation by the hydraulic torques (III) and the direct and cross-coupled stiffness due to angular excitation by the hydraulic torques (IV). First, focusing on the stiffness coefficients of the first submatrix (I), the direct $K_{XX}$, $K_{YY}$ as well as the cross-coupled stiffness $K_{XY}$, $|K_{YX}|$ decreases with increasing modified Reynolds number. Here, the cross-coupled stiffness $K_{XY}$, $|K_{YX}|$ is proportional to $\propto 1/Re_\varphi^*$. Second, focusing on the stiffness coefficients of submatrix (II), the direct stiffness $K_{X \alpha}$, $K_{Y \beta}$ is almost independent on the modified Reynolds number. However, similar to the coefficients of submatrix (II), the cross-coupled stiffness is decreasing with $K_{X \beta}$, $|K_{Y \alpha}|\propto 1/Re_\varphi^*$, cf. submatrix (I). Focusing on the stiffness of submatrix (III), the direct stiffness coefficients $\KAX$, $\KBY$ are almost independent of the modified Reynolds number. The cross-coupled stiffness $K_{\alpha Y}$, $|K_{\beta X}|$ decrease with increasing modified Reynolds number. Finally, the stiffness coefficients of submatrix (IV) are examined. Here, the direct stiffness coefficients $K_{\alpha \alpha}$, $K_{\beta\beta}$ are independent of the modified Reynolds number, whereas the cross-coupled stiffness $K_{\alpha \beta}$, $|K_{\beta\alpha}|$ decrease with increasing modified Reynolds number.\\

Similar to the consideration of the influence of the annulus length on the relevance of the tilt and torque coefficients figure \ref{fig:figure_results_psiRePhi_influenceTilt} shows the influence of the modified Reynolds number on the ratio of tilt to translational stiffness coefficients, cf. equation $\ref{eqn:results_effective_stiffness}$. It exhibits the influence the modified Reynolds number has on the ratio of tilt to translational stiffness coefficients. This reinforces the statement that a sole dependence on the annulus length is insufficient. It can be shown that relevance of the tilt and torque coefficients is proportional $\propto 1/{Re_\varphi^*}^{1.8}$. Therefore, the additional rotordynamic coefficients become more relevant with decreasing modified Reynolds number, i.e. a decreasing gap clearance $\psi$ and Reynolds number $Re_\varphi$.\\

Figure \ref{fig:figure_results_psiRePhi_damping_all} shows the influence of the modified Reynolds number on the direct and cross-coupled damping due to translational excitation by the hydraulic forces (I), the direct and cross-coupled damping due to angular excitation by the hydraulic forces (II), the direct and cross-coupled damping due to translational excitation by the hydraulic torques (III) and the direct and cross-coupled damping due to angular excitation by the hydraulic torques (IV). 
First, focusing on the damping coefficients of submatrix (I), it exhibits decreasing direct coefficients with increasing modified Reynolds number. Similar to the direct stiffness, the direct damping coefficients $C_{XX}$, $C_{YY}$ are proportional $\propto1/Re_\varphi^*$. However, the cross-coupled damping coefficients $C_{XY}$, $C_{YX}$ are almost independent of the modified Reynolds number. Second, focusing on the damping coefficients of submatrix (II) the direct and cross-coupled coefficients do not depend on the modified Reynolds number. In contrast to that, the damping coefficients of submatrix (III) show a slight dependence on the modified Reynolds number. Here, the direct coefficients $C_{\alpha X}$, $C_{\beta Y}$ as well as the cross-coupled coefficients $C_{\alpha Y}$, $|C_{\beta X}|$ decreases slightly with increasing modified Reynolds number. Finally, the damping coefficients of submatrix (IV) are examined. It is shown that the cross-coupled damping $C_{\alpha \beta}$, $|C_{\beta\alpha}|$ slightly increases with increasing modified Reynolds number. In accordance with the direct damping coefficients of submatrix (I), the direct damping $C_{\alpha\alpha}$, $C_{\beta\beta}$ decreases proportionally $\propto 1/Re_\varphi^*$, whereas the cross-coupled damping $C_{\alpha \beta}$, $|C_{\beta\alpha}|$ show a linear dependence on the modified Reynolds number.\\

Figure \ref{fig:figure_results_psiRePhi_inertia_all} shows the influence of the modified Reynolds number on the direct and cross-coupled inertia due to translational excitation by the hydraulic forces (I), the direct and cross-coupled inertia due to angular excitation by the hydraulic forces (II), the direct and cross-coupled inertia due to translational excitation by the hydraulic torques (III) and the direct and cross-coupled inertia due to angular excitation by the hydraulic torques (IV). 
First, focusing on the inertia coefficients of the first submatrix (I), it exhibits slightly increasing inertia coefficients $M_{XX}$, $M_{YY}$ with increasing modified Reynolds number. The cross-coupled inertia coefficients $M_{XY}$, $|M_{YX}|$ are one to two order of magnitudes smaller than the direct ones, exhibiting a slight increase with increasing modified Reynolds number. Second, focusing on the inertia coefficients of submatrix (II) the direct and cross-coupled inertia are almost independent of the modified Rey\-nolds number. It is noted that the inertia coefficients of submatrix (II) are two orders of magnitude smaller than the ones of submatrix (I). Next, considering the inertia coefficients of submatrix (III), a decreasing direct inertia $M_{\alpha X}$, $M_{\beta Y}$ is exhibited with increasing modified Reynolds number. In contrast to that, the cross-coupled inertia $\MAY$, $|\MBX|$ increases with an increasing modified Reynolds number. Finally, focusing on the inertia coefficients of submatrix (IV), the direct inertia coefficients $\MAA$, $\MBB$ show a parabolic behaviour, whereas the cross-coupled inertia $|M_{\alpha \beta}|$, $M_{\beta \alpha}$ linearly increases with increasing modified Reynolds number.\\

In summary, the following statements can be made:
\begin{itemize}
    \item The cross-coupled stiffness and direct damping of submatrix (I) as well as the cross-coupled stiffness of submatrix (II) and the direct damping of submatrix (IV) decrease proportionally $\propto 1/Re_\varphi^*$.
    \item The remaining coefficients only show a marginal influence of the modified Reynolds number.
    \item The relevance of the tilt and torque coefficients decreases proportionally $\propto 1/{Re_\varphi^*}^{1.8}$
\end{itemize}

\subsection{Influence of the flow number}
\begin{figure*}
	\centering
	\includegraphics[scale=0.87]{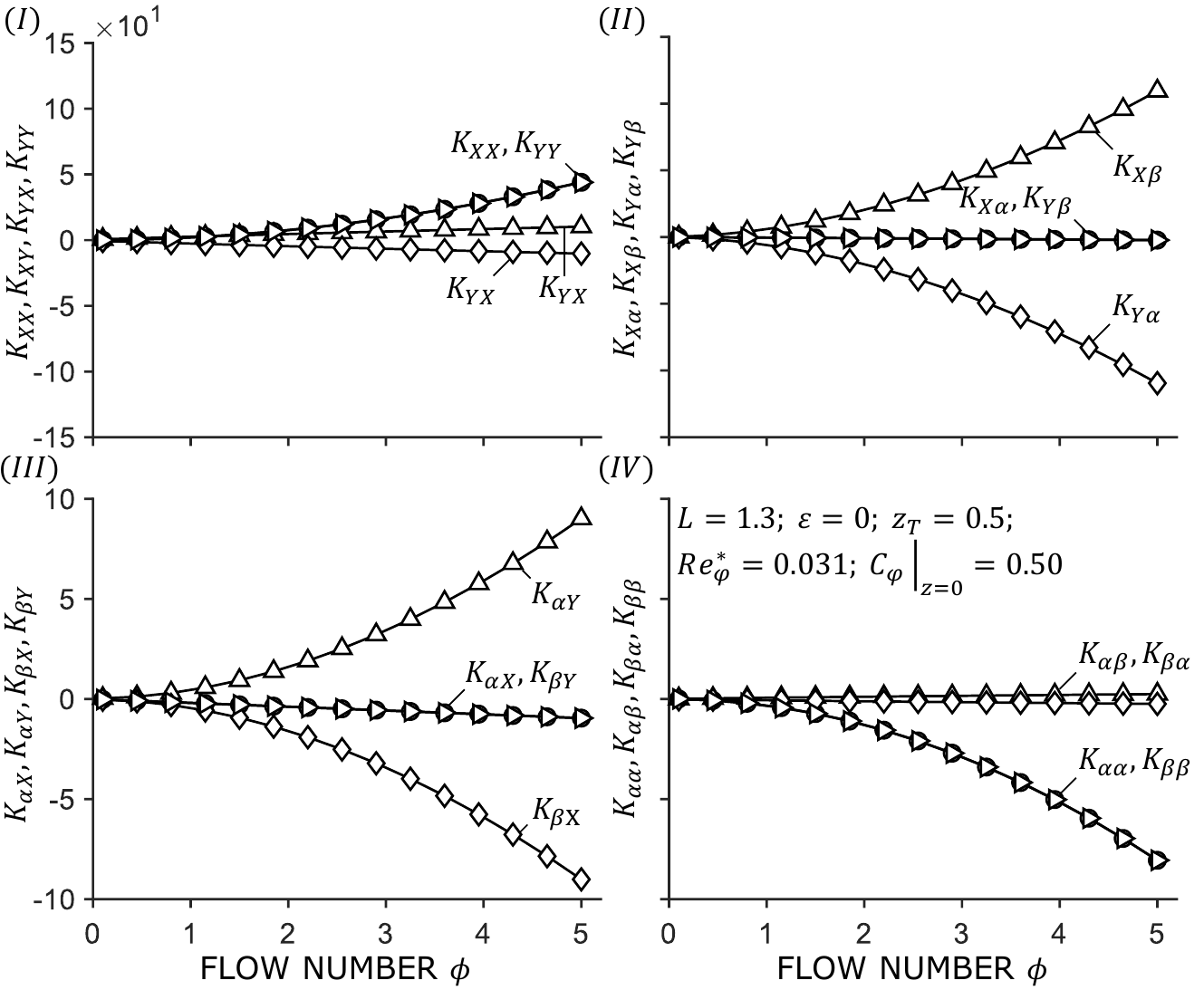}
	\caption{Influence of the flow number on the stiffness due to translational and angular excitation. (I) Stiffness due to translational excitation by the hydraulic forces. (II) Stiffness due to angular excitation by the hydraulic forces. (III) Stiffness due to translational excitation by the hydraulic torques. (IV) Stiffness due to angular excitation by the hydraulic torques.}
	\label{fig:figure_results_phi_stiffness_all}
\end{figure*}

\begin{figure}
	\centering
	\includegraphics[scale=0.87]{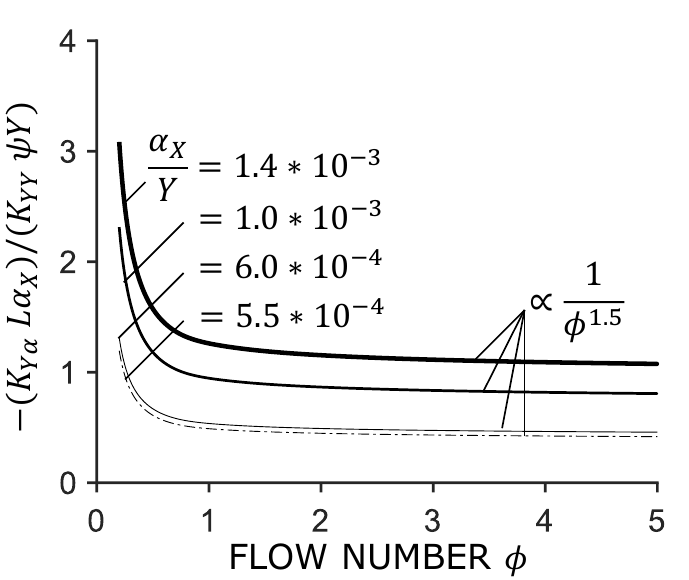}
	\caption{Influence of the flow number and the ratio of angular to translational excitation $\alpha_X/Y$ on the ratio of tilt to translational stiffness coefficients.}
	\label{fig:figure_results_phi_influenceTilt}
\end{figure}

\begin{figure*}
	\centering
	\includegraphics[scale=0.87]{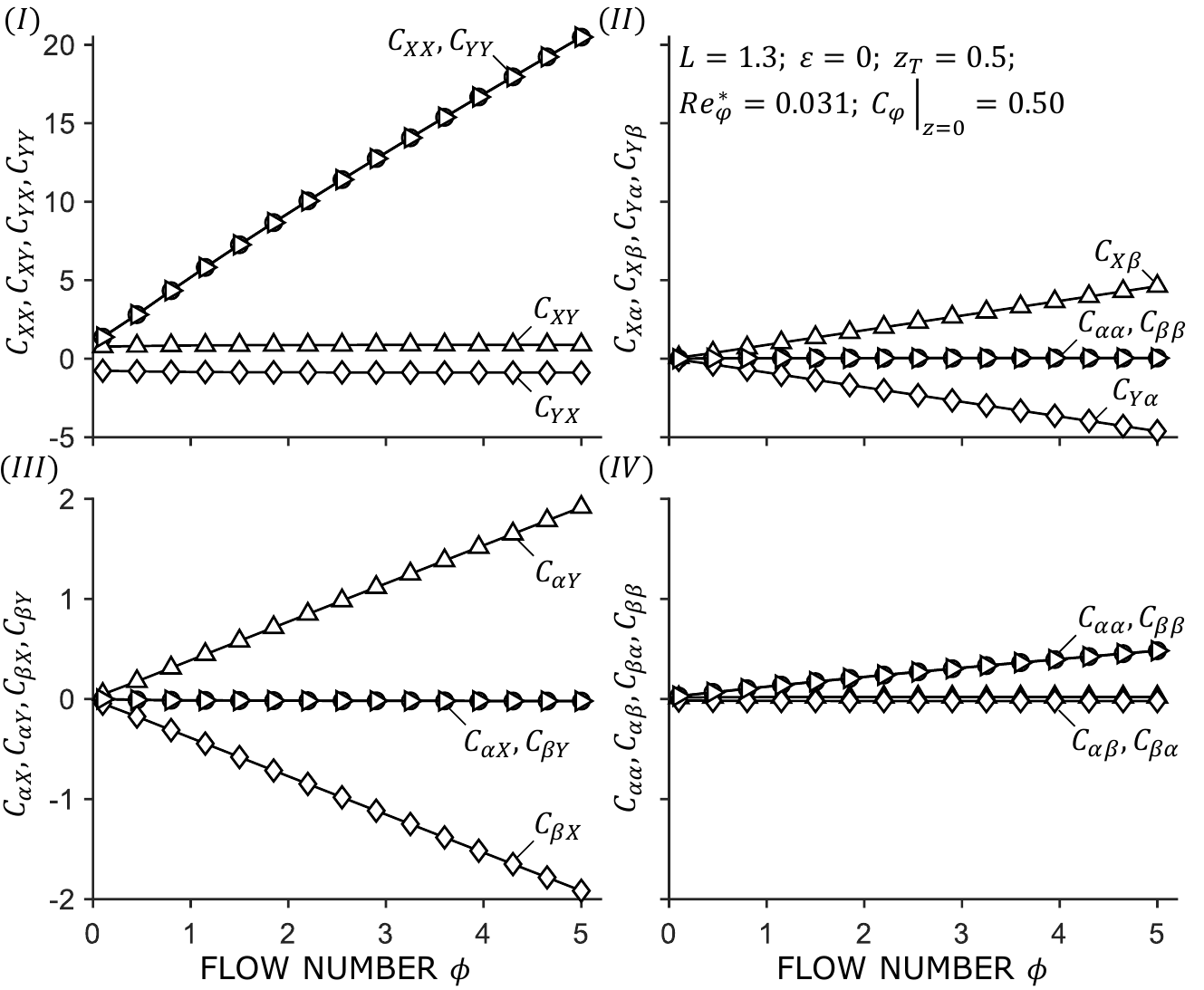}
	\caption{Influence of the flow number on the damping due to translational and angular excitation. (I) Damping due to translational excitation by the hydraulic forces. (II) Damping due to angular excitation by the hydraulic forces. (III) Damping due to translational excitation by the hydraulic torques. (IV) Damping due to angular excitation by the hydraulic torques.}
	\label{fig:figure_results_phi_damping_all}
\end{figure*}

\begin{figure*}
	\centering
	\includegraphics[scale=0.87]{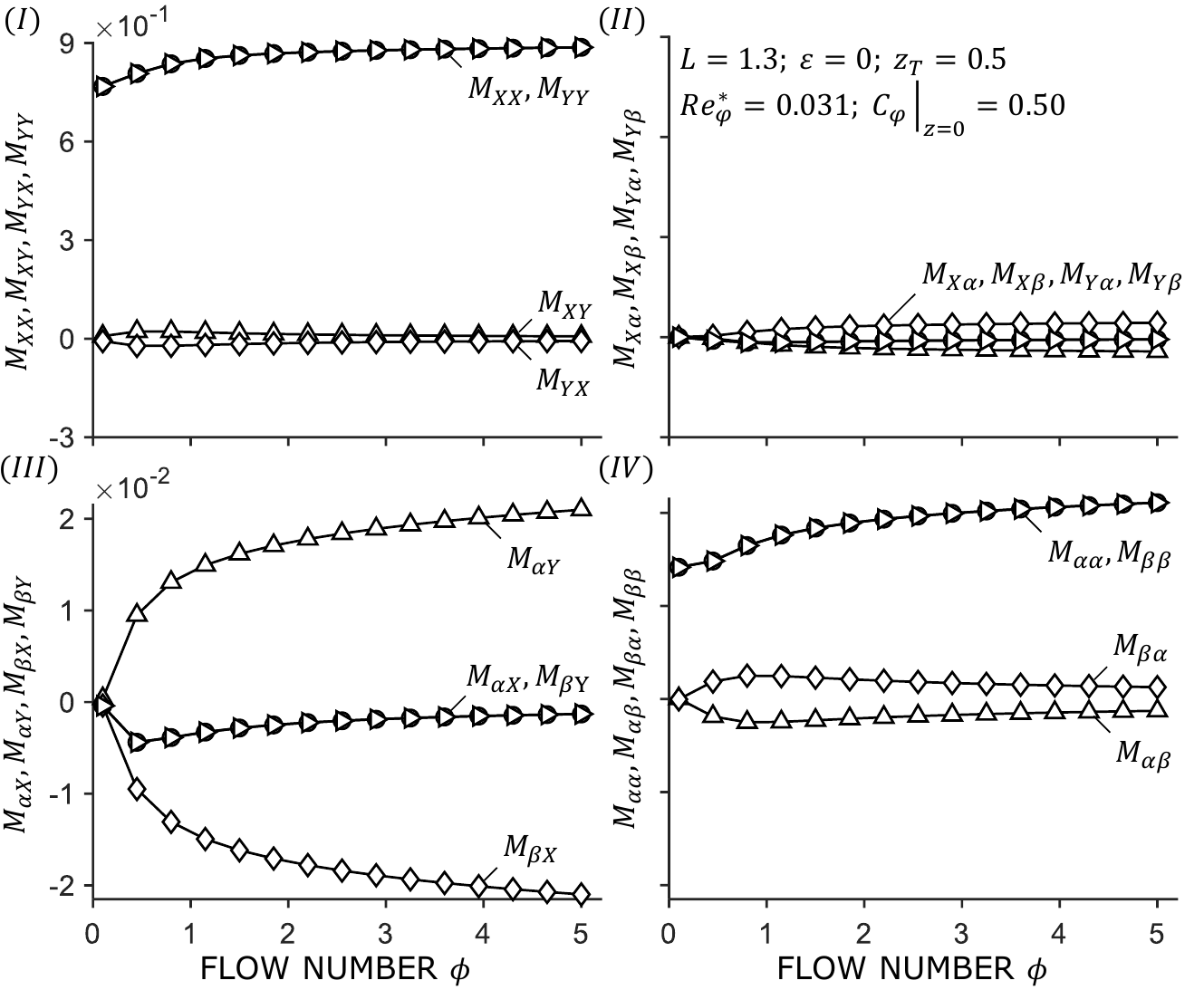}
	\caption{Influence of the flow number on the inertia due to translational and angular excitation. (I) Inertia due to translational excitation by the hydraulic forces. (II) Inertia due to angular excitation by the hydraulic forces. (III) Inertia due to translational excitation by the hydraulic torques. (IV) Inertia due to angular excitation by the hydraulic torques.}
	\label{fig:figure_results_phi_inertia_all}
\end{figure*}

Figures \ref{fig:figure_results_phi_stiffness_all} to \ref{fig:figure_results_phi_inertia_all} give the influence of the flow number on the rotordynamic force and torque coefficients for translational and angular excitation. The annulus with length $L=1.3$ is operated at concentric conditions, i.e. $\varepsilon = 0$ with a modified Reynolds number $Re_\varphi = 0.031$ and a pre-swirl before the annulus $C_\varphi|_{z=0}=0.5$ The fulcrum lies in the centre of the annular gap, i.e. $z_T = 0.5$.\\ 

Figure \ref{fig:figure_results_phi_stiffness_all} shows the influence of the flow number on the direct and cross-coupled stiffness due to translational excitation by the hydraulic forces (I), the direct and cross-coupled stiffness due to angular excitation by the hydraulic forces (II), the direct and cross-coupled stiffness due to translational excitation by the hydraulic torques (III) and the direct and cross-coupled stiffness due to angular excitation by the hydraulic torques (IV). First, focusing on the stiffness coefficients of the first submatrix (I), it exhibits direct stiffness coefficients $K_{XX}$, $K_{YY}$ proportionally increasing with $\propto \phi^{1.9}$, whereas the cross-coupled stiffness $K_{XY}$, $|K_{YX}|$ only increases proportionally $\propto \phi^{0.9}$. The difference in the exponent is due to an increase in flow number, resulting in an increased Lomakin effect. This effect mainly affects the direct stiffness due to the altered axial pressure field. Second, focusing on the stiffness coefficients of submatrix (II), the direct stiffness coefficients $K_{X \alpha}$, $K_{Y \beta}$ are almost independent of the flow number compared to the cross-coupled coefficients. Similar to the direct stiffness of submatrix (I), the cross-coupled stiffness $K_{X \beta}$, $|K_{Y \alpha}|$ increases proportionally $\propto \phi^{1.9}$. It is noted that the cross-coupled stiffness coefficients of submatrix (II) are of the same order of magnitude as the direct coefficients of submatrix (I). Again, this is of particular interest when evaluating the  relevance of the tilt and torque coefficients. Focusing on the stiffness of submatrix (III), the direct stiffness coefficients slightly decrease with increasing flow number, whereas the cross-coupled stiffness $K_{\alpha Y}$, $|K_{\beta X}|$ increases proportionally $\propto \phi^{1.8}$. Finally, focusing on the stiffness coefficients of submatrix (IV), the cross-coupled stiffness coefficients are almost independent of the flow number. The direct stiffness $K_{\alpha \alpha}$, $K_{\beta\beta}$, however, decreases proportionally $\propto -\phi^2$.\\
Similar to the previous consideration of the tilt an torque coefficients, figure \ref{fig:figure_results_phi_influenceTilt} shows the influence of the flow number on the ratio of tilt to translational stiffness coefficients, cf. equation $\ref{eqn:results_effective_stiffness}$. In contrast to the influence regarding the annulus length, the influence of the flow number decreases with increasing flow number. Here, the ratio of tilt to translational stiffness coefficients is proportional $\propto 1/\phi^{1.5}$. Therefore, the additional rotordynamic coefficients become more relevant with decreasing flow number, i.e. a decreasing axial flow component. This is  due to the fact that the direct stiffness of submatrix (I) $K_{YY}$ decreases faster than the cross-coupled stiffness $K_{Y \alpha}$, becoming $K_{YY}=0$ at vanishing flow component, i.e. $\phi=0$.\\

Figure \ref{fig:figure_results_phi_damping_all} shows the influence of the flow number on the direct and cross-coupled damping due to translational excitation by the hydraulic forces (I), the direct and cross-coupled damping due to angular excitation by the hydraulic forces (II), the direct and cross-coupled damping due to translational excitation by the hydraulic torques (III) and the direct and cross-coupled damping due to angular excitation by the hydraulic torques (IV).\\ 

First, focusing on the damping coefficients of the first submatrix (I), it exhibits the values of cross-coupled damping coefficients $C_{XY}$, $|C_{YX}|$ being independent of the flow number. However, the direct damping $C_{XX}$, $C_{YY}$ shows a linear dependence on the flow number. Second, focusing on the damping coefficients of submatrix (II) the direct damping coefficients $\CXA$, $\CYB$ are almost independent of the flow number, whereas the cross-coupled damping $\CXB$, $|\CYA|$ linearly increases with the flow number. The same dependency can be seen when considering the damping of submatrix (III). Being in the same order of magnitude as the damping of submatrix (II), the direct damping coefficients $C_{\alpha X}$, $C_{\beta Y}$ are almost independent of the flow number, whereas the cross-coupled damping $|C_{\beta X}|$, $C_{\alpha Y}$ increases linearly with the flow number. Finally, focusing on the damping of submatrix (IV) the cross-coupled damping coefficients $|C_{\alpha \beta}|$, $C_{\beta\alpha}$ are independent of the flow number, whereas the direct damping $C_{\alpha\alpha}$, $C_{\beta\beta}$ increases linearly.\\

Figure \ref{fig:figure_results_phi_inertia_all} shows the influence of the flow number on the direct and cross-coupled inertia due to translational excitation by the hydraulic forces (I), the direct and cross-coupled inertia due to angular excitation by the hydraulic forces (II), the direct and cross-coupled inertia due to translational excitation by the hydraulic torques (III) and the direct and cross-coupled inertia due to angular excitation by the hydraulic torques (IV). \\
First, focusing on the inertia coefficients of the first submatrix (I), it exhibits inertia coefficients almost independent of the flow number. Merely the direct inertia $M_{XX}$, $M_{YY}$ shows an influence at low flow numbers. Here, the direct inertia increases with increasing flow number. This is due to the fact that the inertia coefficients originate from a displacement of the fluid inside the annulus. Here, the axial flow component is almost negligible, resulting only in small changes due to an increasing flow number. Second, focusing on the inertia coefficients of submatrix (II), the direct coefficients $\MXA$, $\MYB$ are independent of the flow number, whereas the cross-coupled coefficients $\MXB$, $|\MYA|$ slightly increase. It is noted that the inertia coefficients of submatrix (II) are one to two orders of magnitude smaller than the ones of submatrix (I). Focusing on the inertia of submatrix (III), the cross-coupled inertia coefficients exhibit a dependence o the flow number. Here, the coefficients $\MAY$, $|\MBX|$ increase with increasing flow number, whereas the direct inertia changed insignificantly. Finally, focusing on the inertia of submatrix (IV) the coefficients behave in a similar manner as the coefficients of submatrix (I). Here, the direct inertia $\MAA$, $\MBB$ increases with increasing flow number, whereas the cross-coupled coefficients $|M_{\alpha \beta}|$, $M_{\beta \alpha}$ increase at first, reaching a maximum at $\phi = 0.8$. By further increasing the flow number the cross-coupled inertia coefficients decrease.\\

In summary, the following statements can be made:
\begin{itemize}
    \item The direct stiffness of submatrix (I) as well as the cross-coupled stiffness of submatrix (II) and (IV) increase proportionally $\propto \phi^{1.9}$.
    \item The damping coefficients exhibit a linear dependence on the flow number.
    \item The relevance of the tilt and torque coefficients decreases proportionally $\propto 1/\phi^{1.5}$.
\end{itemize}

\section{Conclusions}
In the presented paper we discuss the dynamic force and torque characteristic of annular gaps with an axial flow component. First, the rotordynamic influence of annular gaps is discussed. So far there is a severe lack of understanding with regard to the dynamic characteristic including hydraulic forces and torques of the flow inside the annulus. Second, a new calculation method is presented, using a perturbed integro-differential approach in combination with power-law ansatz functions and a Hirs' model to calculate the dynamic force and torque characteristics. For validation purposes, the Clearance-Averaged Pressure Model (CAPM) is compared to existing literature by \cite{Nordmann.1988, Nelson.1988,Simon.1992}, exhibiting a good agreement with the results shown therein. Third, an extensive parameter study is carried out. The results are used to evaluate the relevance of the tilt and torque coefficients. It is shown that the preconceived idea of an overall threshold depending only on the annulus length is insufficient. Rather, the operating conditions of the turbomachinery has to be taken into account. The influence of the operating conditions is particularly evident when considering the modified Reynolds number, the flow number and the ratio of the excitation amplitudes in translational and rotational degree of freedom.  

\section*{Acknowledgements}
We gratefully acknowledge the financial support of the Federal Ministry for Economic Affairs and Energy (BMWi) due to an enactment of the German Bundestag under Grant No. 03EE5036B and KSB SE \& Co. KGaA. In addition, we gratefully acknowledge the financial support of the industrial collective research programme (IGF no. 21029 N/1), supported by the Federal Ministry for Economic Affairs and Energy (BMWi) through the AiF (German Federation of Industrial Research Associations e.V.) due to an enactment of the German Bundestag. Special gratitude is expressed to the participating companies and their representatives in the accompanying industrial committee for their advisory and technical support.

\section*{Declaration of competing interest}
The authors declare that they have no known personal relationships or competing financial interests that could have appeared to influence the work reported in this paper.

\bibliographystyle{asmems4}
\bibliography{main-refs}

\newpage
\appendix\label{sec:appendix}
\clearpage
\section{Influence of the eccentricity}
\begin{figure*}
	\centering
	\includegraphics[scale=0.87]{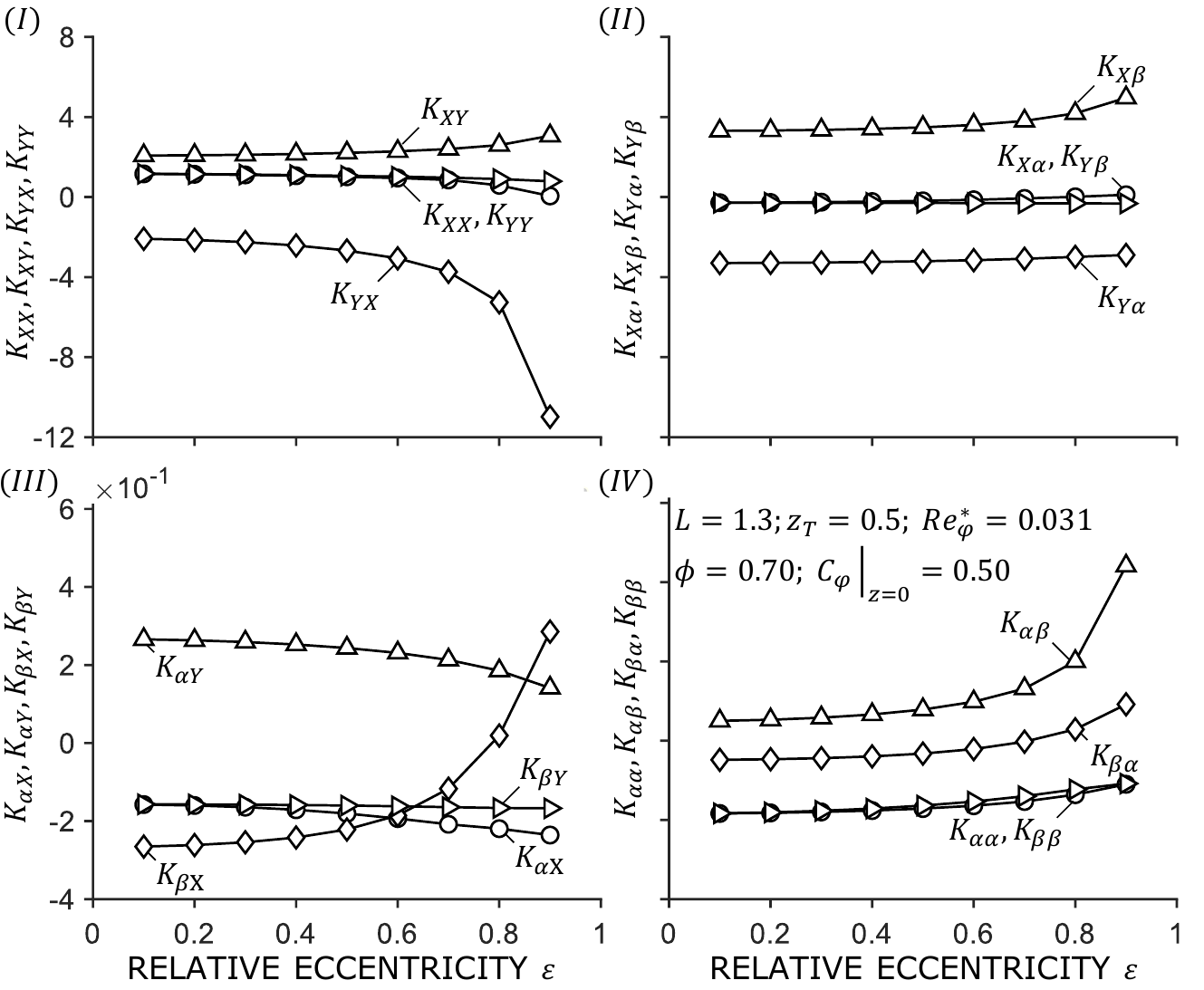}
	\caption{Influence of eccentricity on the stiffness due to translational and angular excitation. (I) Stiffness due to translational excitation by the hydraulic forces. (II) Stiffness due to angular excitation by the hydraulic forces. (III) Stiffness due to translational excitation by the hydraulic torques. (IV) Stiffness due to angular excitation by the hydraulic torques.}
	\label{fig:figure_results_eps_stiffness_all}
\end{figure*}

\begin{figure}
	\centering
	\includegraphics[scale=0.87]{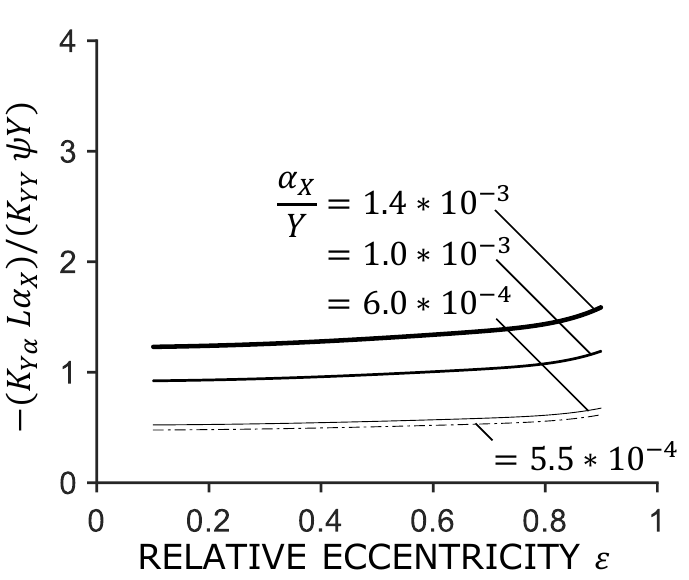}
	\caption{Influence of the eccentricity and the ratio of angular to translational excitation $\alpha_X/Y$ on the ratio of tilt to translational stiffness coefficients.}
	\label{fig:figure_results_eps_influenceTilt}
\end{figure}

\begin{figure*}
	\centering
	\includegraphics[scale=0.87]{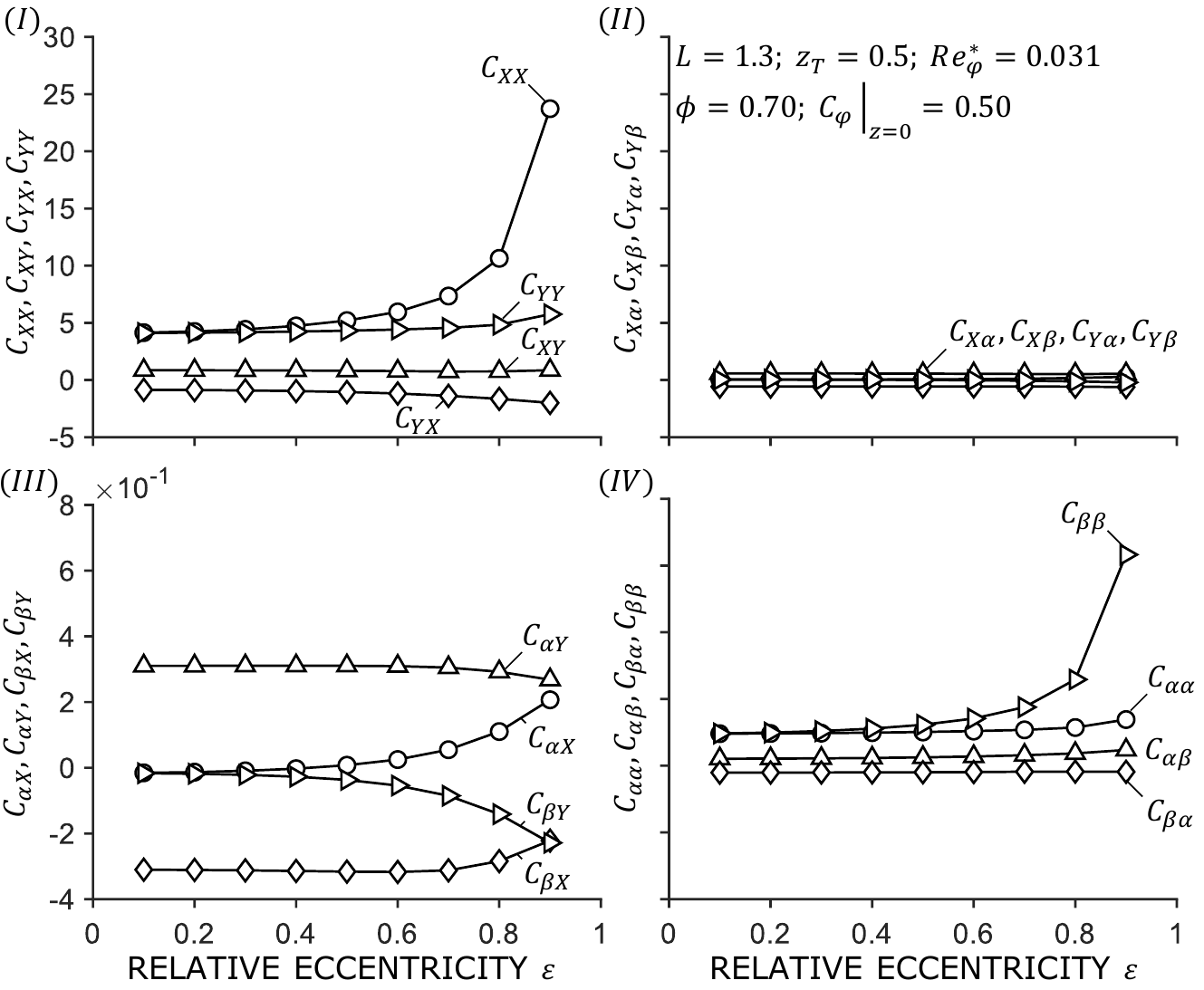}
	\caption{Influence of eccentricity on the damping due to translational and angular excitation. (I) Damping due to translational excitation by the hydraulic forces. (II) Damping due to angular excitation by the hydraulic forces. (III) Damping due to translational excitation by the hydraulic torques. (IV) Damping due to angular excitation by the hydraulic torques.}
	\label{fig:figure_results_eps_damping_all}
\end{figure*}

\begin{figure*}
	\centering
	\includegraphics[scale=0.87]{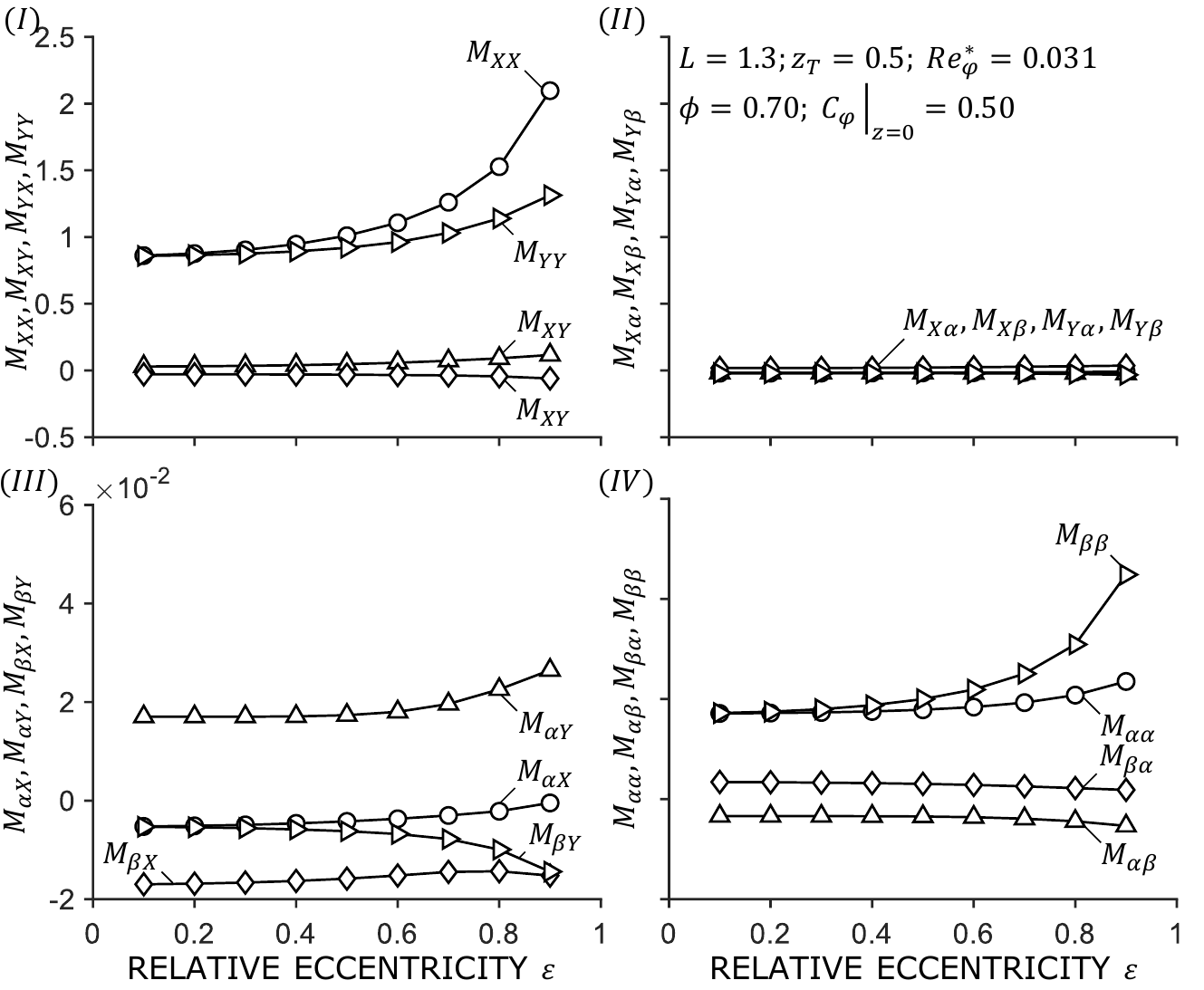}
	\caption{Influence of eccentricity on the inertia due to translational and angular excitation. (I) Inertia due to translational excitation by the hydraulic forces. (II) Inertia due to angular excitation by the hydraulic forces. (III) Inertia due to translational excitation by the hydraulic torques. (IV) Inertia due to angular excitation by the hydraulic torques.}
	\label{fig:figure_results_eps_inertia_all}
\end{figure*}

In the following, the influence of the eccentricity on the rotordynamic coefficients is investigated. Figures \ref{fig:figure_results_eps_stiffness_all} to \ref{fig:figure_results_eps_inertia_all} give the force and torque coefficients for translational and angular excitation. The annulus of length $L = 1.3$ is operated at a modified Reynolds number $Re_\varphi = 0.031$ and flow number $\phi = 0.7$. The pre-swirl before the annulus is set to $C_\varphi|_{z=0}=0.5$ and the fulcrum lies in the centre of the gap, i.e. $z_T = 0.5$. In contrast to the skew-symmetric sub-matrices when investigating concentric operation conditions, i.e. $\varepsilon = 0$, the skew-symmetry of the sub-matrices vanishes with increasing eccentricity.\\

Figure \ref{fig:figure_results_eps_stiffness_all} shows the influence of an eccentric operated shaft on the direct and cross-coupled stiffness due to translational excitation by the hydraulic forces (I), the direct and cross-coupled stiffness due to angular excitation by the hydraulic forces (II), the direct and cross-coupled stiffness due to translational excitation by the hydraulic torques (III) and the direct and cross-coupled stiffness due to angular excitation by the hydraulic torques (IV). First, focusing on the stiffness coefficients of the first submatrix (I), it exhibits a skew-symmetric submatrix up to an eccentricity $\varepsilon \approx 0.4$. This is in good agreement with the existing literature, cf. \cite{Childs.1993}. Furthermore, the curve of the direct and cross-coupled stiffness becomes non-linear when increasing the eccentricity. Here, the direct stiffness $K_{XX}$, $K_{YY}$ slightly decreases with increasing eccentricity, whereas the cross-coupled stiffness $K_{XY}$ slightly increases. In contrast, the cross-coupled stiffness $K_{YX}$ decreases exponentially with increasing eccentricity. Second, focusing on the stiffness coefficients of submatrix (II) the direct and cross-coupled stiffness are almost independent of the eccentricity. Here, the direct stiffness $K_{X\alpha}$ slightly decreases, whereas the direct stiffness $K_{Y\beta}$ slightly increases with increasing eccentricity. Furthermore, the cross-coupled stiffness $K_{Y\alpha}$ decreases, whereas the cross-coupled stiffness $\KXB$ increases with increasing eccentricity. It is noted that the stiffness coefficients of submatrix (II) are in the same order of magnitude as the ones of submatrix (I). In contrast to the coefficients of submatrix (II) the stiffness coefficients of submatrix (III) exhibit an eccentricity influence. Here, the direct stiffness $\KAX$, $\KBY$ as well as the cross-coupled stiffness $\KAY$ decrease with increasing eccentricity. However, the cross-coupled stiffness $\KBX$ shows an exponential progression and a sign change at $\varepsilon \approx 0.75$ when increasing the eccentricity. Finally, focusing on the direct and cross-coupled stiffness of submatrix (IV) the direct stiffness $\KAA$, $\KBB$ as well as the cross-coupled stiffness $\KBA$ decrease with increasing eccentricity, whereas the cross-coupled stiffness $\KAB$ increases exponentially with increasing eccentricity.\\

Similar to the consideration of the influence of length on the relevance of the tilt and torque coefficients, figure \ref{fig:figure_results_eps_influenceTilt} shows the influence of the eccentricity on the ratio of tilt to translational stiffness coefficients, cf. equation $\ref{eqn:results_effective_stiffness}$. It is shown that, although the additional coefficients are relevant at the selected operating point for an annulus length $L=1.3$, the influence of the eccentricity is negligible.\\
 
Figure \ref{fig:figure_results_eps_stiffness_all} shows the influence of an eccentrically operated rotor on the direct and cross-coupled damping due to translational excitation by the hydraulic forces (I), the direct and cross-coupled damping due to angular excitation by the hydraulic forces (II), the direct and cross-coupled damping due to translational excitation by the hydraulic torques (III) and the direct and cross-coupled damping due to angular excitation by the hydraulic torques (IV). 
First, focusing on the damping coefficients of the first submatrix (I), it exhibits a skew-symetric submatrix up to an eccentricity $\varepsilon = 0.4$. This is in good agreement with the corresponding stiffness coefficients, cf. figure \ref{fig:figure_results_eps_stiffness_all}. Furthermore, the curve of the direct and cross-coupled damping becomes non-linear when increasing the eccentricity. Here, the direct damping $\CXX$ increases exponentially when increasing the eccentricity, whereas $\CYY$ only increases at high eccentricities. The cross-coupled damping coefficients are almost independent of the eccentricity, with $\CYX$ slightly decreasing with increasing eccentricity. Second, focusing on the damping coefficients of submatrix (II) the direct and cross-coupled damping coefficients are almost independent of the eccentricity, being one order of magnitude smaller than the coefficients of submatrix (I). In contrast to the coefficients of submatrix (II) the damping coefficients of submatrix (III) exhibit an eccentricity influence. Here, the direct and cross-coupled damping coefficients $\CAX$, $\CBX$ increase with increasing eccentricity, whereas the direct and cross-coupled damping coefficients $\CBY$, $\CAY$ decrease. Finally, focusing on the direct and cross-coupled damping of submatrix (IV), the direct damping $\CBB$ exponentially increases with with increasing eccentricity, whereas the direct and cross-coupled damping coefficients $\CAA$, $\CAB$, $\CBA$ are almost independent of the eccentricity.\\

Figure \ref{fig:figure_results_eps_stiffness_all} shows the influence of an eccentrically operated rotor on the direct and cross-coupled inertia due to translational excitation by the hydraulic forces (I), the direct and cross-coupled inertia due to angular excitation by the hydraulic forces (II), the direct and cross-coupled inertia due to translational excitation by the hydraulic torques (III) and the direct and cross-coupled inertia due to angular excitation by the hydraulic torques (IV). 
First, focusing on the inertia coefficients of the first submatrix (I), it exhibits a skew-symmetric submatrix up to an eccentricity $\varepsilon = 0.4$. Furthermore, the direct inertia exponentially increases with increasing eccentricity. Here, the direct inertia $\MXX$ increases faster than the  direct inertia $\MYY$. The cross-coupled inertia coefficients are almost independent of the eccentricity, with $\MXY$ slightly increasing and $\MYX$ slightly decreasing with increasing eccentricity. Second, focusing on the inertia coefficients of submatrix (II) the direct and cross-coupled inertia coefficients are almost independent of the eccentricity, being two orders of magnitude smaller than the coefficients of submatrix (I). In contrast to the coefficients of submatrix (II), the damping coefficients of submatrix (III) exhibit an eccentricity influence. Here, the direct and cross-coupled damping coefficients $\MAX$, $\MAY$ and $\MBX$ increase with increasing eccentricity, whereas the direct inertia $\MBY$ decrease with increasing eccentricity. Finally, focusing on the direct and cross-coupled inertia of submatrix (IV), the direct inertia behaves similar to the direct inertia of submatrix (I). $\MBB$ exponentially increases with with increasing eccentricity. The cross-coupled inertia coefficients $\MAB$, $\MBA$ slightly decrease with increasing eccentricity.\\

\section{Influence of the centre of rotation}
\begin{figure*}
	\centering
	\includegraphics[scale=0.87]{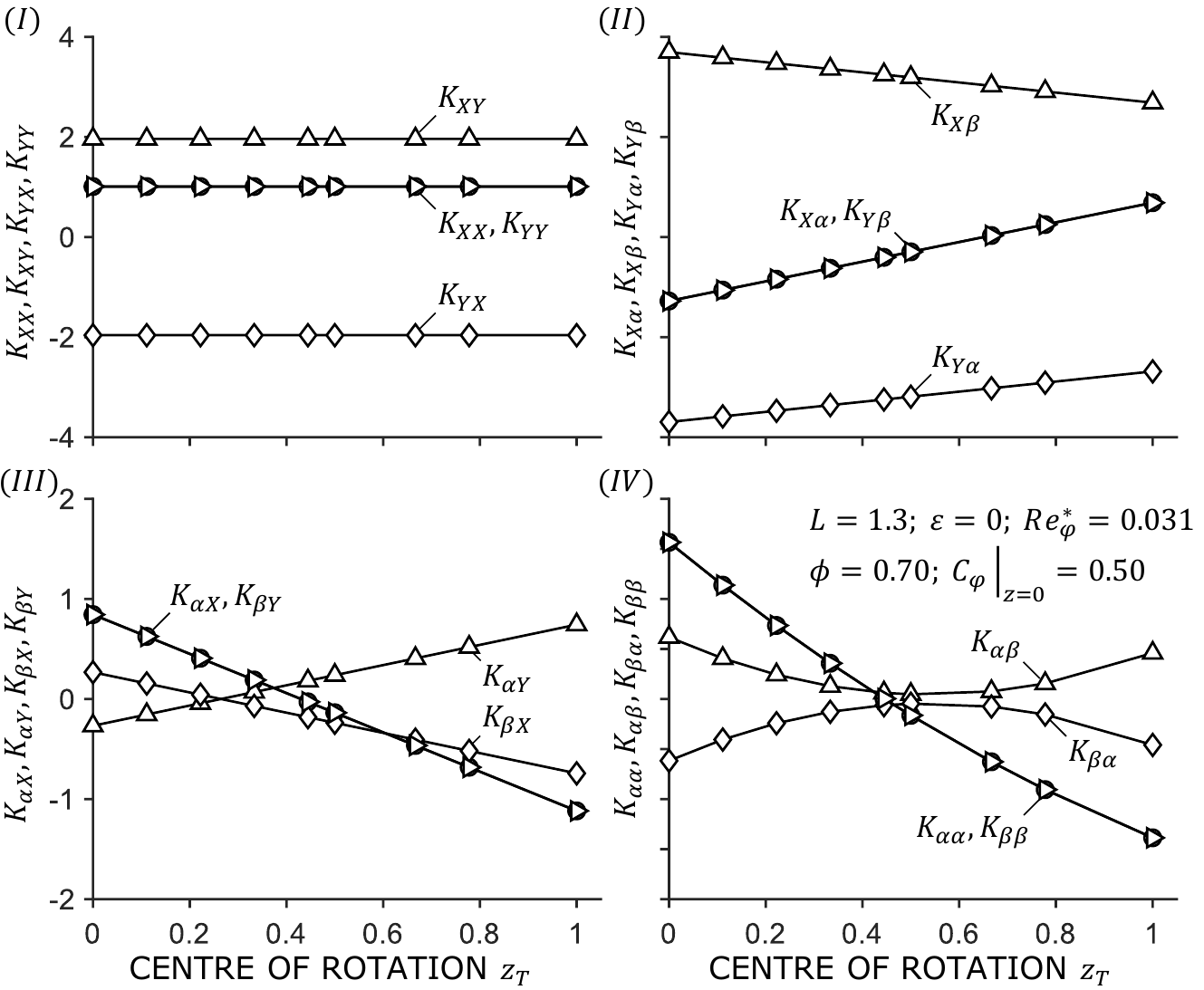}
	\caption{Influence of the centre of rotation on the stiffness due to translational and angular excitation. (I) Stiffness due to translational excitation by the hydraulic forces. (II) Stiffness due to angular excitation by the hydraulic forces. (III) Stiffness due to translational excitation by the hydraulic torques. (IV) Stiffness due to angular excitation by the hydraulic torques.}
	\label{fig:figure_results_fulcrum_stiffness_all}
\end{figure*}

\begin{figure}
	\centering
	\includegraphics[scale=0.87]{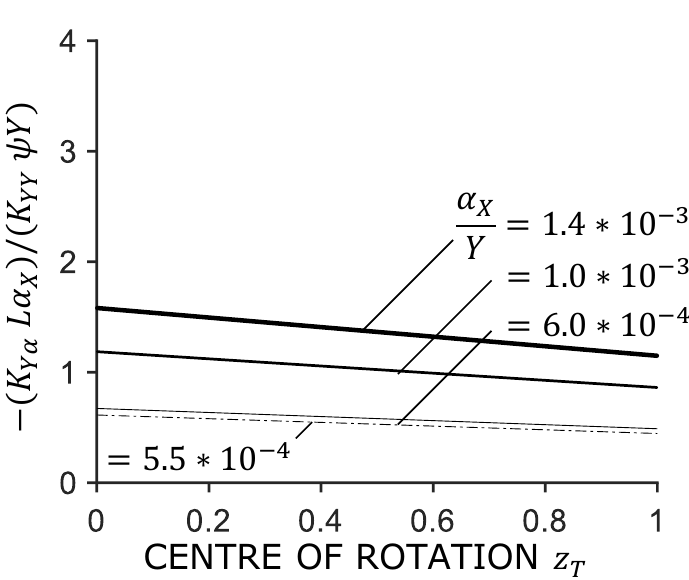}
	\caption{Influence of the centre of rotation and the ratio of angular to translational excitation $\alpha_X/Y$ on the ratio of tilt to translational stiffness coefficients.}
	\label{fig:figure_results_fulcrum_influenceTilt}
\end{figure}

\begin{figure*}
	\centering
	\includegraphics[scale=0.87]{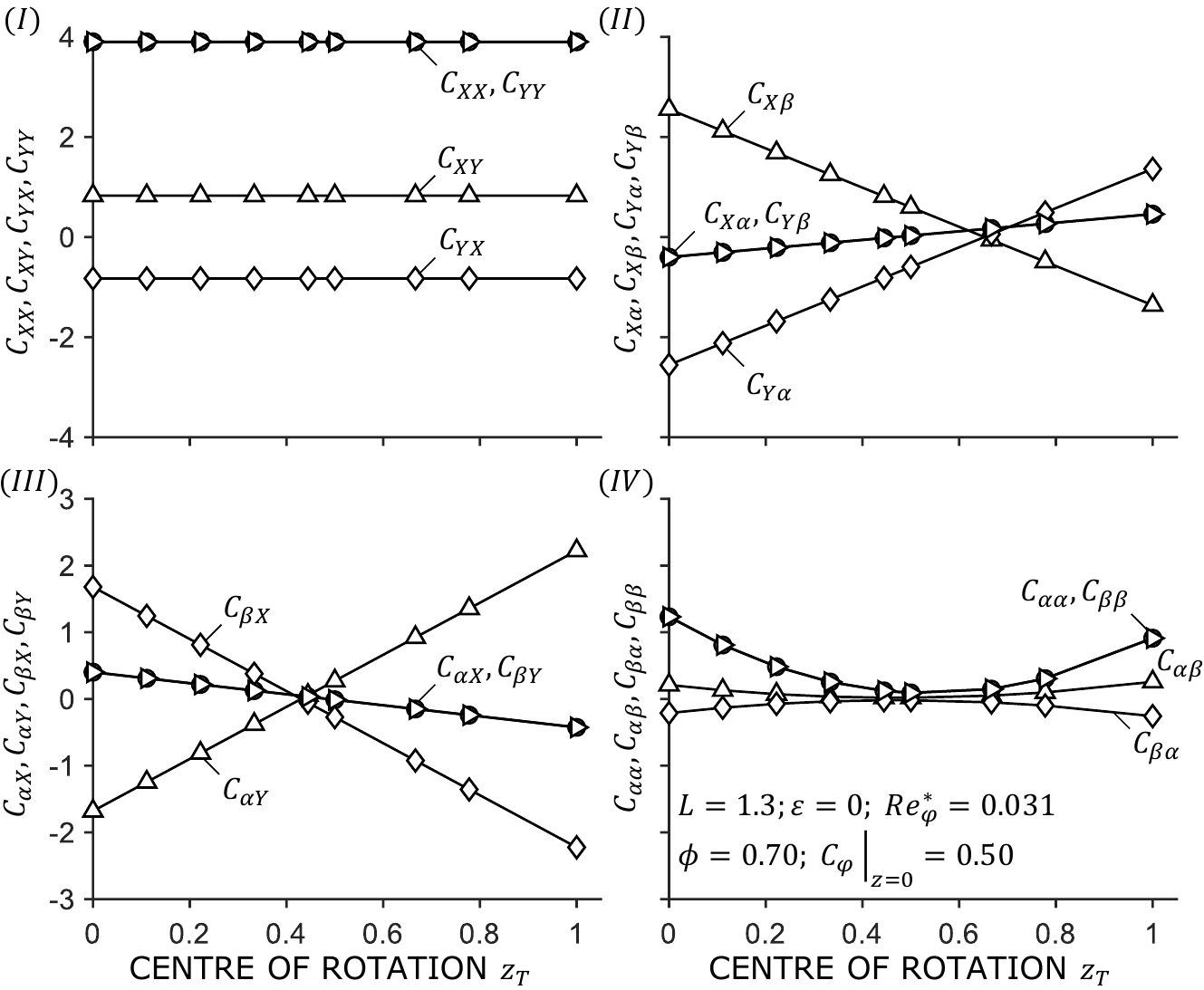}
	\caption{Influence of the centre of rotation on the damping due to translational and angular excitation. (I) Damping due to translational excitation by the hydraulic forces. (II) Damping due to angular excitation by the hydraulic forces. (III) Damping due to translational excitation by the hydraulic torques. (IV) Damping due to angular excitation by the hydraulic torques.}
	\label{fig:figure_results_fulcrum_damping_all}
\end{figure*}

\begin{figure*}
	\centering
	\includegraphics[scale=0.87]{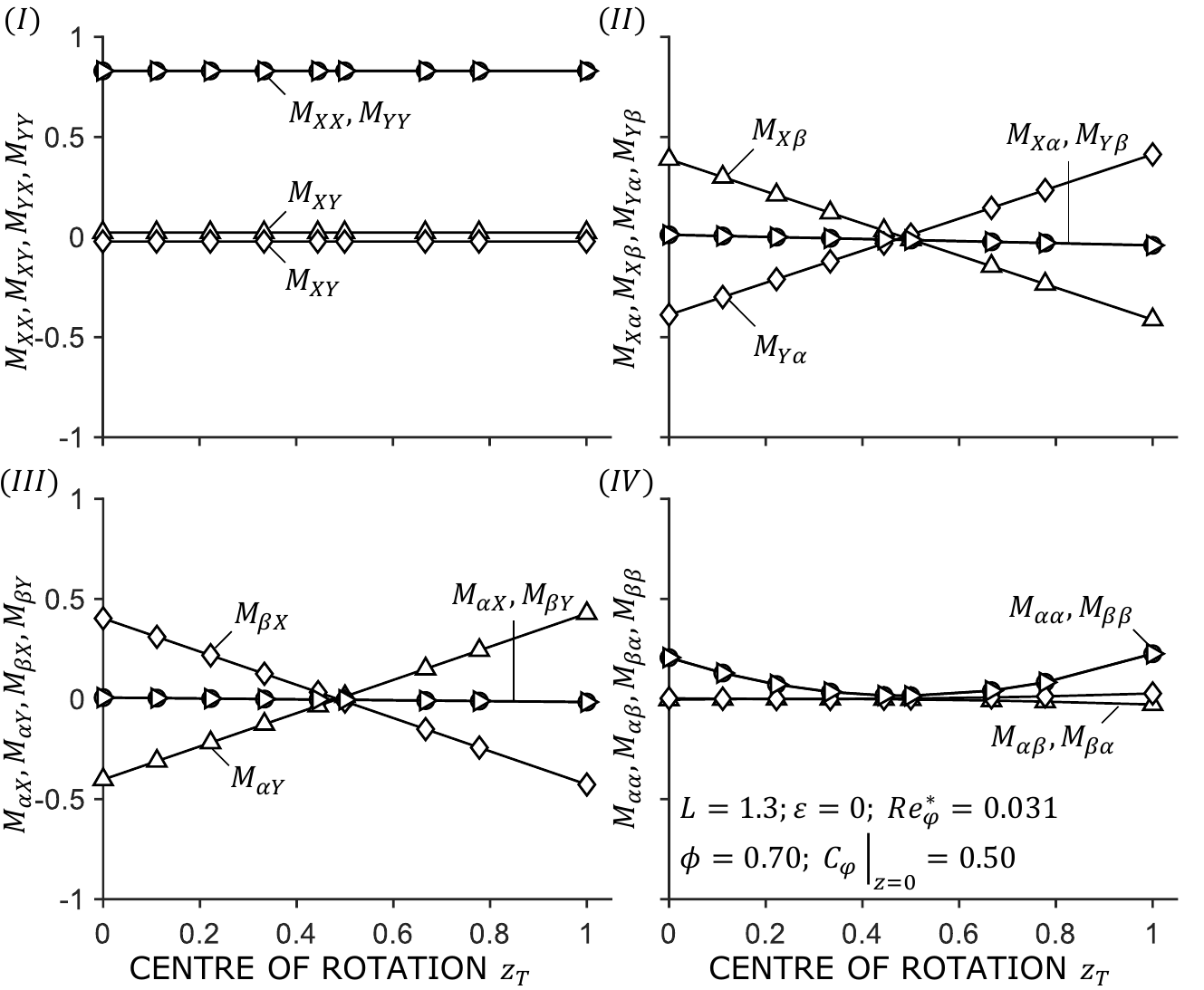}
	\caption{Influence of the centre of rotation on the inertia due to translational and angular excitation. (I) Inertia due to translational excitation by the hydraulic forces. (II) Inertia due to angular excitation by the hydraulic forces. (III) Inertia due to translational excitation by the hydraulic torques. (IV) Inertia due to angular excitation by the hydraulic torques.}
	\label{fig:figure_results_fulcrum_inertia_all}
\end{figure*}

In the following, the influence of the centre of rotation on the rotordynamic coefficients is investigated. Figures \ref{fig:figure_results_fulcrum_stiffness_all} to \ref{fig:figure_results_fulcrum_inertia_all} give the force and torque coefficients for translational and angular excitation. The annulus of length $L = 1.3$ is operated at concentric conditions, i.e. $\varepsilon = 0$ with a modified Reynolds number $Re_\varphi = 0.031$ and flow number $\phi = 0.7$. The pre-swirl before the annulus is set to $C_\varphi|_{z=0}=0.5$. Due to the concentric operation conditions, the matrices are skew-symmetric.\\

Figure \ref{fig:figure_results_fulcrum_stiffness_all} shows the influence of the centre of rotation on the direct and cross-coupled stiffness due to translational excitation by the hydraulic forces (I), the direct and cross-coupled stiffness due to angular excitation by the hydraulic forces (II), the direct and cross-coupled stiffness due to translational excitation by the hydraulic torques (III) and the direct and cross-coupled stiffness due to angular excitation by the hydraulic torques (IV). First, focusing on the stiffness coefficients of the first submatrix (I), it exhibits coefficients that are independent of the centre of rotation. This is due to the fact that the coefficients originate from the translational excitation by the hydraulic forces. As will be shown, all coefficients of submatrix (I), i.e. stiffness $K_I$, damping $C_I$ and inertia $M_I$, are independent for the centre of rotation. Second, focusing on the stiffness coefficients of submatrix (II) the direct and cross-coupled coefficients show a linear dependence on the centre of rotation. Here, the direct stiffness coefficients $\KXA$, $\KYB$ increase with the centre of rotation moving form the inlet to the outlet of the annulus, whereas the cross-coupled stiffness coefficients $\KXB$, $|\KYA|$ decrease with moving centre of rotation. It is noted that the stiffness coefficients of submatrix (II) are of the same order of magnitude as the ones of submatrix (I). Focusing on the stiffness coefficients of submatrix (III), they exhibit a linear dependence on the centre of rotation. Here, the direct stiffness $\KAX$, $\KBY$ and the cross-coupled stiffness $\KBX$ decrease with moving centre of rotation, whereas the cross-coupled stiffness $\KAY$ increases linearly with the centre of rotation moving form the inlet to the outlet of the annulus. In contrast to the linear dependence on the centre of rotation of submatrices (II) and (III), the stiffness of submatrix (IV) show a parabolic behaviour. Here, the direct stiffness $\KAA$, $\KBB$ strongly decreases with the centre of rotation moving form the annulus inlet to the annulus outlet. Furthermore, the cross-coupled stiffness $\KAB$, $|\KBA|$ first decreases with moving centre of rotation, reaching a minimum at $z_T= 0.54$. By moving the centre of rotation further toward the outlet of the annulus, the cross-coupled stiffness increases. The fact that the minimum does not coincide with a centre of rotation $z_T= 0.5$ originated from the axial flow component. Due to that, a bias torque is induced on the rotor, shifting the minimum towards the outlet of the annulus.\\

Similar to the consideration of the influence of length on the relevance of the tilt and torque coefficients, figure \ref{fig:figure_results_fulcrum_influenceTilt} shows the influence of the centre of rotation on the ratio of tilt to translational stiffness coefficients, cf. equation $\ref{eqn:results_effective_stiffness}$. It is shown that although the additional coefficients are relevant at the selected operating point for an annulus length $L=1.3$, the influence of centre of rotation on the effective stiffness is negligible.\\
 
Figure \ref{fig:figure_results_fulcrum_damping_all} shows the influence of the centre of rotation on the direct and cross-coupled damping due to translational excitation by the hydraulic forces (I), the direct and cross-coupled damping due to angular excitation by the hydraulic forces (II), the direct and cross-coupled damping due to translational excitation by the hydraulic torques (III) and the direct and cross-coupled damping due to angular excitation by the hydraulic torques (IV). 
First, focusing on the damping coefficients of the first submatrix (I), it exhibits damping coefficients independent of the centre of rotation. Second, focusing on the damping coefficients of submatrix (II), the direct and cross-coupled damping show a linear dependence on the centre of rotation. Here, the direct damping coefficients $\CXA$, $\CYB$ increase with the centre of rotation moving from the inlet to the outlet of the annulus, whereas the cross-coupled damping $\CXB$, $|\CYA|$ decreases with moving centre of rotation. Focusing on the damping coefficients of submatrix (III), the coefficients also exhibit a linear dependence on the centre of rotation. Here, the direct damping $\CAX$, $\CBY$ and the cross-coupled damping $\CBX$ decrease with moving centre of rotation, whereas the cross-coupled damping $\CAY$ increases linearly with the centre of rotation moving from the inlet to the outlet of the annulus. Similar to the stiffness coefficients of submatrix (IV) the damping coefficients also exhibit a parabolic dependence on the centre of rotation. Here, the direct damping coefficients $\CAA$, $\CBB$ as well as the cross-coupled damping $\CAB$, $|\CBA|$ decrease at first, reaching a minimum at $z_T = 0.54$ for the direct damping and $z_T = 0.47$ for the cross-coupled damping. By further moving the the centre of rotation towards the outlet of the annulus, the corresponding coefficients increase.\\

Figure \ref{fig:figure_results_fulcrum_inertia_all} shows the influence of the centre of rotation on the direct and cross-coupled inertia due to translational excitation by the hydraulic forces (I), the direct and cross-coupled inertia due to angular excitation by the hydraulic forces (II), the direct and cross-coupled inertia due to translational excitation by the hydraulic torques (III) and the direct and cross-coupled inertia due to angular excitation by the hydraulic torques (IV). First, focusing on the inertia coefficients of the first submatrix (I), it exhibits inertia coefficients independent of the centre of rotation, similar to the stiffness and damping coefficients. Second, focusing on the inertia coefficients of submatrix (II), the direct inertia coefficients $\MXA$, $\MYB$ are independent of the centre of rotation, whereas the cross-coupled coefficients $\MXB$, $\MYA$ show a linear dependency. Here, the cross-coupled inertia $\MXB$ decreases with the centre of rotation while the cross-coupled inertia $\MYA$ increases. Focusing on the inertia coefficients of submatrix (III), the cross-coupled coefficients exhibit a linear dependence on the centre of rotation. Similar to the direct inertia coefficients of submatrix (II) the ones of submatrix (III) are also independent of the centre of rotation. Here, the cross-coupled inertia $\MBX$ decreases with moving centre of rotation, whereas the cross-coupled inertia $\MAY$ increases linearly with the centre of rotation moving from the inlet to the outlet of the annulus. In accordance with the stiffness and damping coefficients of submatrix (IV), the inertia coefficients show a parabolic behaviour. Here, the cross-coupled inertia $|\MAB|$, $\MBA$ slightly increases with the centre of rotation moving from the inlet to the outlet of the annulus, whereas the direct inertia $\MAA$, $\MBB$ decreases at first, reaching a minimum at $z_T= 0.49$. By moving the centre of rotation further towards the outlet of the annulus, the direct inertia increases again.\\

\section{Influence of the pre-swirl}
\begin{figure*}
	\centering
	\includegraphics[scale=0.87]{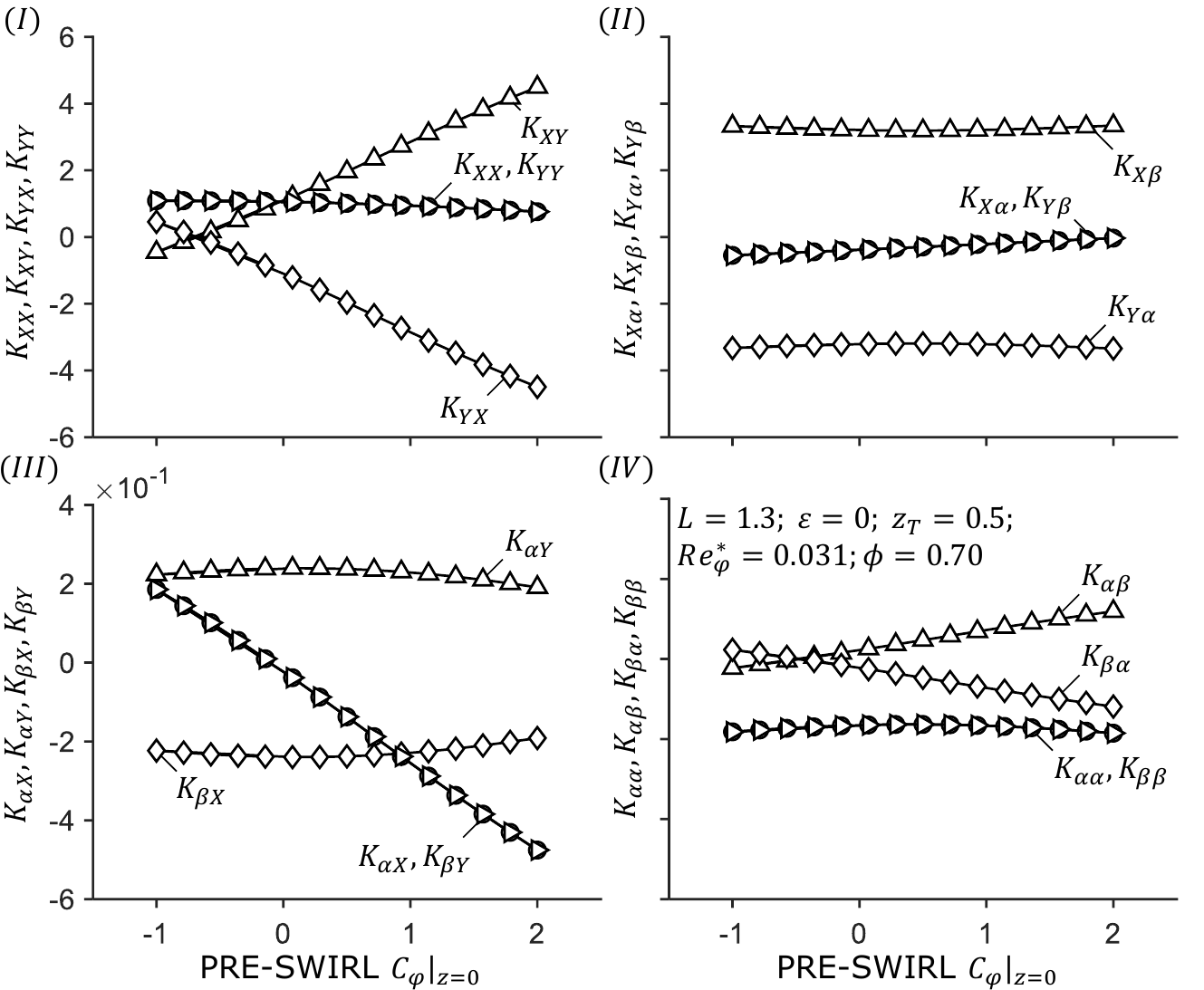}
	\caption{Influence of the pre-swirl on the stiffness due to translational and angular excitation. (I) Stiffness due to translational excitation by the hydraulic forces. (II) Stiffness due to angular excitation by the hydraulic forces. (III) Stiffness due to translational excitation by the hydraulic torques. (IV) Stiffness due to angular excitation by the hydraulic torques.}
	\label{fig:figure_results_pre-swirl_stiffness_all}
\end{figure*}

\begin{figure}
	\centering
	\includegraphics[scale=0.87]{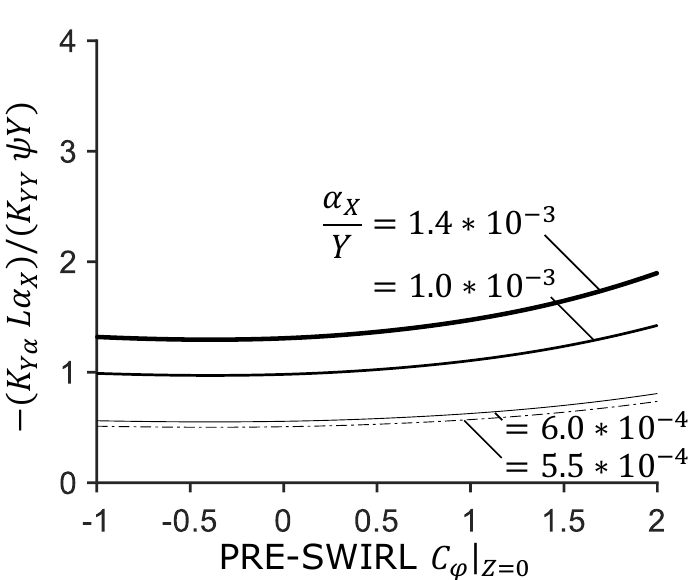}
	\caption{Influence of the pre-swirl and the ratio of angular to translational excitation $\alpha_X/Y$ on the ratio of tilt to translational stiffness coefficients.}
	\label{fig:figure_results_pre-swirl_influenceTilt}
\end{figure}

\begin{figure*}
	\centering
	\includegraphics[scale=0.87]{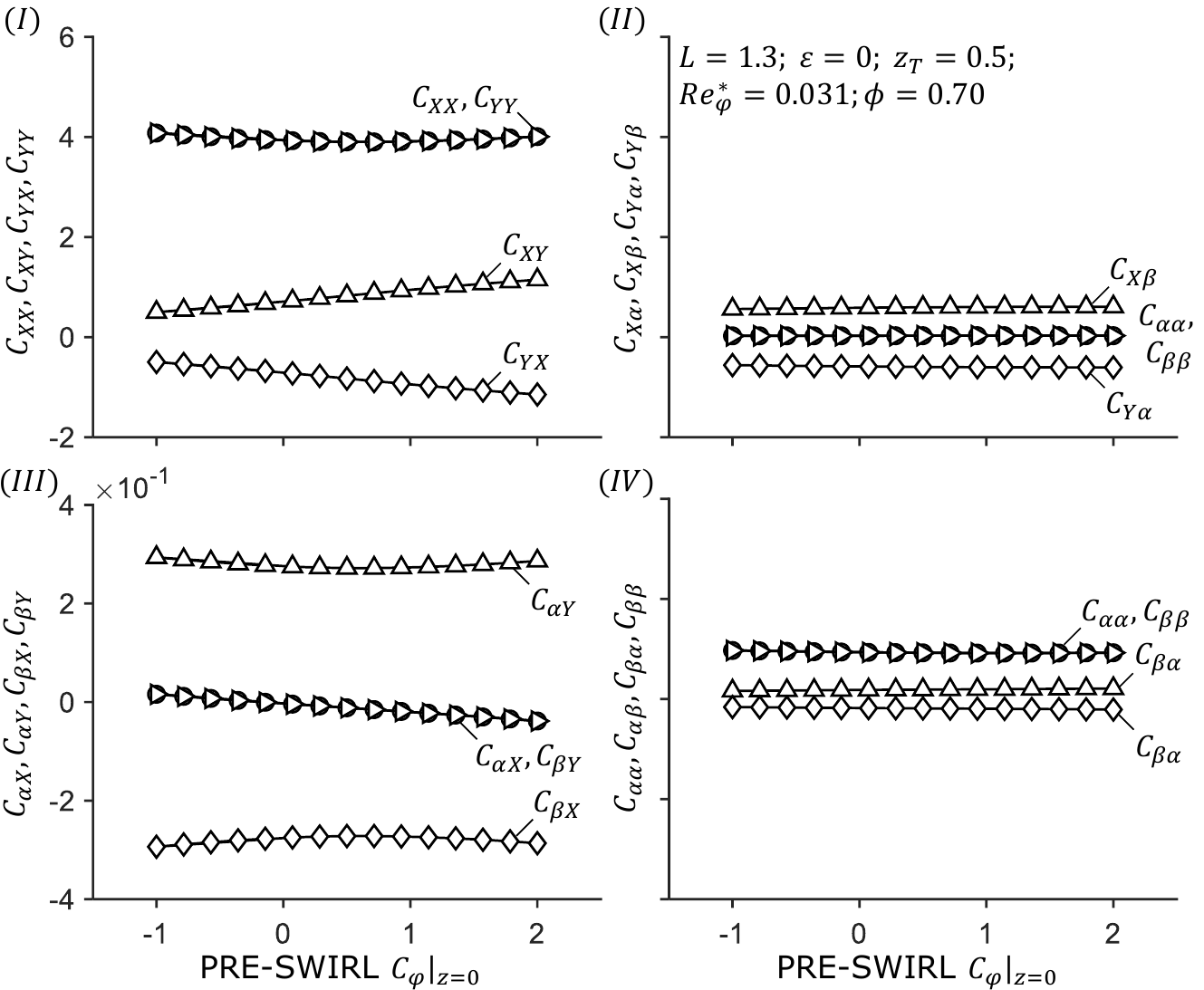}
	\caption{Influence of the pre-swirl on the damping due to translational and angular excitation. (I) Damping due to translational excitation by the hydraulic forces. (II) Damping due to angular excitation by the hydraulic forces. (III) Damping due to translational excitation by the hydraulic torques. (IV) Damping due to angular excitation by the hydraulic torques.}
	\label{fig:figure_results_pre-swirl_damping_all}
\end{figure*}

\begin{figure*}
	\centering
	\includegraphics[scale=0.87]{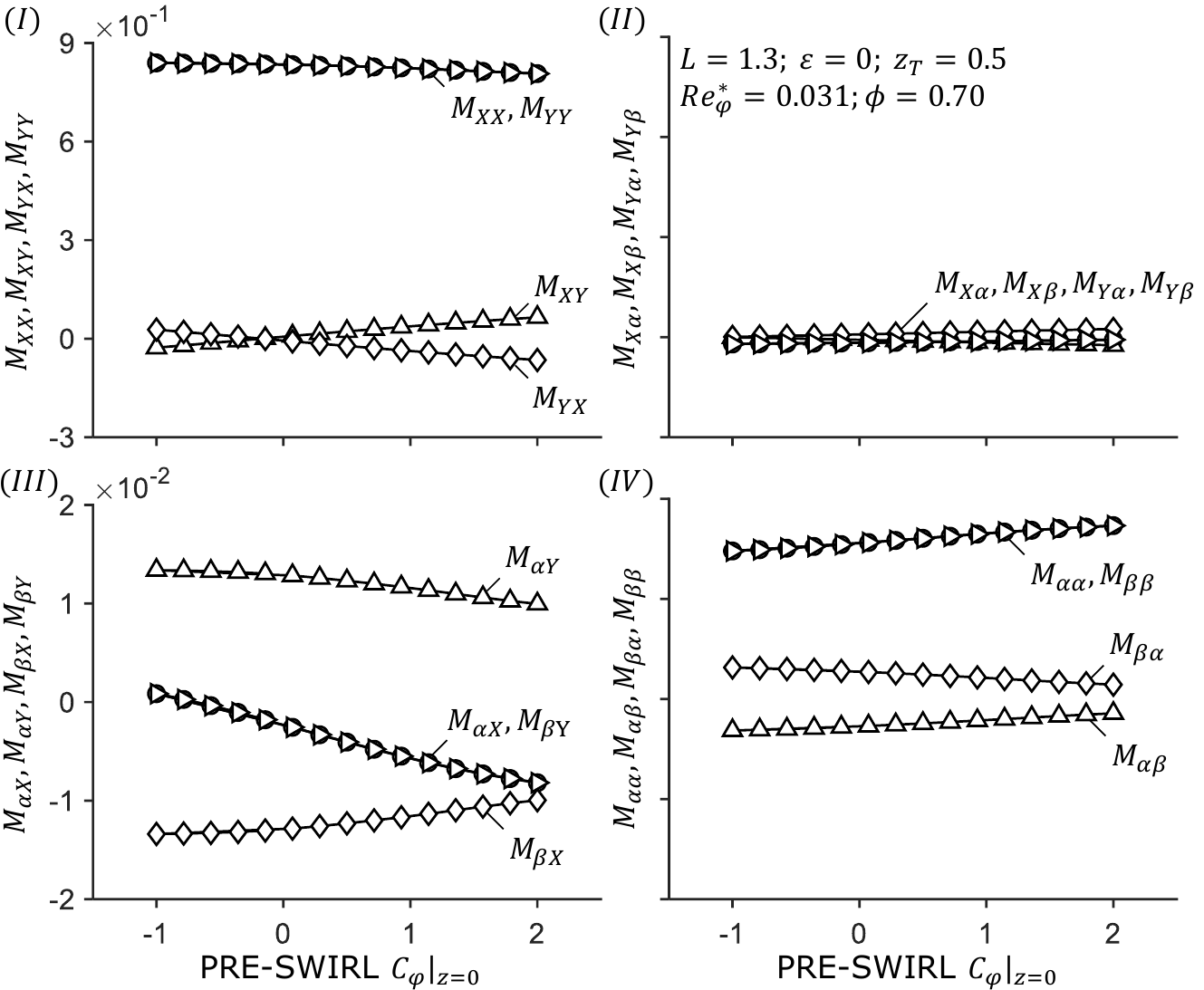}
	\caption{Influence of the pre-swirl on the inertia due to translational and angular excitation. (I) Inertia due to translational excitation by the hydraulic forces. (II) Inertia due to angular excitation by the hydraulic forces. (III) Inertia due to translational excitation by the hydraulic torques. (IV) Inertia due to angular excitation by the hydraulic torques.}
	\label{fig:figure_results_pre-swirl_inertia_all}
\end{figure*}

Figures \ref{fig:figure_results_pre-swirl_stiffness_all} to \ref{fig:figure_results_pre-swirl_inertia_all} give the influence of pre-swirl on the rotordynamic force and torque coefficients for translational and angular excitation. The annulus of length $L = 1.3$ is operated at concentric conditions, i.e. $\varepsilon = 0$ with a modified Reynolds number $Re_\varphi = 0.031$ and flow number $\phi = 0.7$. The pre-swirl before the annulus is set to $C_\varphi|_{z=0}=0.5$ and the fulcrum lies in the centre of the annular gap, i.e. $z_T = 0.5$.\\

Figure \ref{fig:figure_results_pre-swirl_stiffness_all} shows the influence of the pre-swirl on the direct and cross-coupled stiffness due to translational excitation by the hydraulic forces (I), the direct and cross-coupled stiffness due to angular excitation by the hydraulic forces (II), the direct and cross-coupled stiffness due to translational excitation by the hydraulic torques (III) and the direct and cross-coupled stiffness due to angular excitation by the hydraulic torques (IV).
First, focusing on the stiffness coefficients of the first submatrix (I), it exhibits direct stiffness coefficients $\KXX$, $\KYY$ independent of the pre-swirl, whereas the cross-coupled stiffness coefficients $\KXY$, $|\KYX|$ increase linearly with an increasing pre-swirl. This is due to the fact that an increased pre-swirl mainly alters the tangential force component, i.e. $F_Y$, on the rotor. It is noted that the cross-coupled stiffness is of particular interest when analysing the stability of the system. Here, high pre-swirl rations act in a destabilising manner. Second, focusing on the stiffness coefficients of submatrix (II), the direct and cross-coupled stiffness are almost independent of the pre-swirl. Only the direct stiffness coefficients slightly increase with an increasing pre-swirl. Focusing on the stiffness of submatrix (III), the cross-coupled stiffness coefficients $\KAY$, $|\KBX|$ sightly decrease with an increasing pre-swirl, whereas the direct stiffness coefficients $\KAX$, $\KBY$ decrease linearly with the pre-swirl ratio. Finally focusing on the stiffness coefficients of submatrix (IV) the direct stiffness $\KAA$, $\KBB$ is almost independent of the pre-swirl. In contrast, the cross-coupled stiffness shows a linear dependence on the pre-swirl.\\

Similar to the previous consideration on the relevance of the $48$ coefficients, figure \ref{fig:figure_results_pre-swirl_influenceTilt} shows the influence of the pre-swirl on the ratio of tilt to translational stiffness coefficients, cf. equation $\ref{eqn:results_effective_stiffness}$. Similar to the influence of the eccentricity and the centre of rotation, the pre-swirl only slightly affects the effective stiffness. Here, the influence only becomes apparent with large pre-swirl ratios, i.e $C_\varphi|_{z=0}$. \\

Figure \ref{fig:figure_results_pre-swirl_damping_all} shows the influence of the pre-swirl on the direct and cross-coupled damping due to translational excitation by the hydraulic forces (I), the direct and cross-coupled damping due to angular excitation by the hydraulic forces (II), the direct and cross-coupled damping due to translational excitation by the hydraulic torques (III) and the direct and cross-coupled damping due to angular excitation by the hydraulic torques (IV). 
First, focusing on the damping coefficients of the first submatrix (I), it exhibits direct damping coefficients $\CXX$, $\CYY$ almost independent of the pre-swirl. However, the cross-coupled damping coefficients $\CXY$, $|\CYX|$ increase linearly with the pre-swirl. Second, focusing on the damping coefficients of submatrix (II), the direct and cross-coupled damping are independent of the pre-swirl. This is not only true for the damping coefficients of submatrix (II) but also holds for the damping coefficients of submatrix (III) and submatrix (IV). Solely the direct damping coefficients of submatrix (III) slightly decrease with increasing pre-swirl.\\

Figure \ref{fig:figure_results_pre-swirl_inertia_all} shows the influence of the pre-swirl on the direct and cross-coupled inertia due to translational excitation by the hydraulic forces (I), the direct and cross-coupled inertia due to angular excitation by the hydraulic forces (II), the direct and cross-coupled inertia due to translational excitation by the hydraulic torques (III) and the direct and cross-coupled inertia due to angular excitation by the hydraulic torques (IV). First, focusing on the inertia coefficients of the first submatrix (I), it exhibits direct inertia coefficients $\MXX$, $\MYY$ independent of the pre-swirl. However, the cross-coupled inertia linearly depends on the pre-swirl. Here the cross-coupled inertia coefficients experience a sign change at negative pre-swirl ratios $C_\varphi|_{z=0}<-0.17$. Second, focusing on the inertia coefficients of submatrix (II), the direct and cross-coupled inertia are independent of the pre-swirl. Focusing on the inertia coefficients of submatrix (III), the direct inertia coefficients $\MAX$, $\MBY$ as well as the cross-coupled coefficient $|\MBX|$, $\MAY$ decrease with increasing pre-swirl. Finally, focusing on the inertia coefficients of submatrix (IV), the direct coefficients $\MAA$, $\MBB$ linearly increase with increasing pre-swirl, whereas the cross-coupled inertia coefficients $|\MAB|$, $\MBA$ slightly decrease with increasing pre-swirl ratio.\\


\end{document}